\documentclass[lettersize,journal]{IEEEtran}
\usepackage{amsmath,amsfonts}
\usepackage{algorithmic}
\usepackage{algorithm}
\usepackage{array}
\usepackage[caption=false,font=normalsize,labelfont=sf,textfont=sf]{subfig}
\usepackage{textcomp}
\usepackage{stfloats}
\usepackage{url}
\usepackage{verbatim}
\usepackage{graphicx}
\usepackage{cite}
\usepackage{booktabs}
\usepackage{adjustbox}
\usepackage{makecell}
\usepackage{multirow}
\usepackage{cite}
\usepackage{amsmath,amssymb,amsfonts}
\usepackage{algorithmic}
\usepackage{graphicx}
\usepackage{textcomp}
\usepackage{xcolor}
\usepackage{hyperref}
\usepackage{colortbl}
\usepackage{graphicx}
\newcommand{\rtposeedit}[1]{\textcolor{black}{#1}}

\begin{document}

\title{Physics-Bounded mmWave Sensing for Schedulable, Privacy-Preserving Human Pose Estimation}

\author{
\IEEEauthorblockN{
Shuntian Zheng, 
Hongyang He, 
Jiaqi Li, 
Xiaoman Lu, 
Doeon Kim, 
Jae-Ho Choi, 
Jin Zeng, 
Shuai He, 
Yu Guan\\
}
\IEEEauthorblockA{
shuntian.zheng,
Hongyang.He,
jiaqi.li.16,
Xiaoman.Lu@warwick.ac.uk,
ilsin205@soongsil.ac.kr,
\\
jhochoi@dgist.ac.kr, 
zengjin@tongji.edu.cn,
hs19951021@bupt.edu.cn,
yu.guan@warwick.ac.uk
}
}

\markboth{}%
{Zheng \MakeLowercase{\textit{et al.}}: Physics-Bounded mmWave Sensing for Schedulable, Privacy-Preserving Human Pose Estimation}


\maketitle

\begin{abstract}
Millimeter-wave (mmWave) is a promising modality for human pose estimation (HPE) in mobile deployments with strong privacy requirements and limited resources, such as fall detection in bathrooms or activity monitoring in bedrooms, where cameras are inadmissible and computationally demanding processing is infeasible. 
Although mmWave signals naturally confine human reflections to compact, physically bounded regions, the algorithmic foundations of existing systems fail to provide deterministic execution and accuracy guarantees. 
They either process the full spectrum uniformly, resulting in unpredictable latency that varies across different scenes, or apply lossy compression that discards vital pose structures.
To address this, we present PRISM, a framework that exploits the spatial concentration of RF reflections to achieve schedulable edge HPE. PRISM introduces three core components: 1) Physics-Bounded Integral Processing (PBIP), which restricts computation via constant-time integral queries; 2) Physics-Adaptive Instance Proposal (PAIP), which decomposes scenes involving multiple people into bounded local subproblems; and 3) Deadline-Aware Operation Profiles (DAOP), which provide offline-verified worst-case bounds for runtime quality-latency trade-offs. \textcolor{black}{We evaluate PRISM on four public datasets spanning diverse radar configurations, reporting physical-bound and pose-accuracy measurements across this suite and examining deadline-aware scheduling on multi-person recordings together with an additional single-person set. Under single-threaded isolated execution, PRISM reduces 99th-percentile latency by 24\%--58\% relative to baselines that miss the deadline, records a 0.0\% miss rate on the evaluated traces, and attains the highest pose accuracy among deadline-feasible configurations, providing a practical route toward schedulable mmWave sensing on mobile edge hardware.} 
Code is available in the supplementary material.
\end{abstract}


\section{Introduction}
\label{sec:intro}
\begin{figure}[t]
  \centering
  \includegraphics[width=\linewidth]{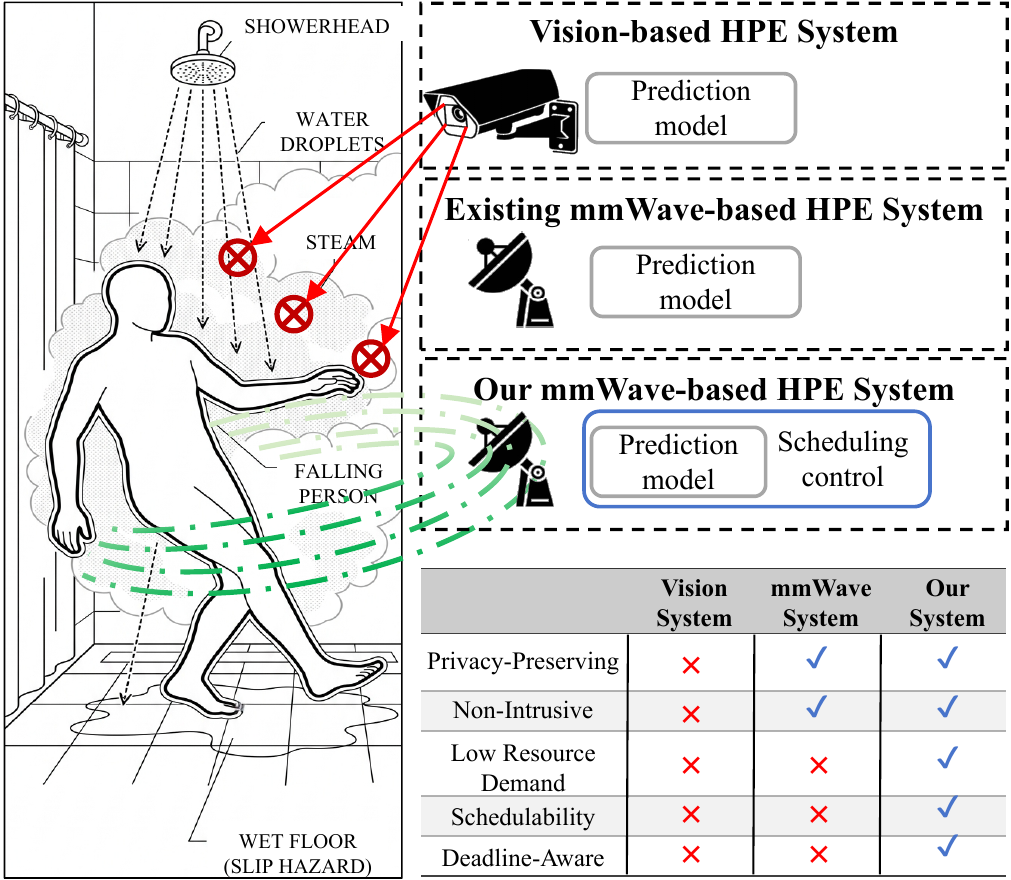}
  \caption{The bathroom is a classic scenario with both privacy sensitivity and computational constraints, where vision-based HPE systems are affected by steam and infringe on privacy, whereas mmWave-based systems offer clear advantages.}
  \label{fig:teaser}
\end{figure}

Human Pose Estimation (HPE), which infers the spatial configuration of body joints from sensor data, is foundational to intelligent mobile edge applications such as fall detection in bathrooms, activity monitoring in bedrooms, and rehabilitation tracking in home care settings \cite{deng2023midas++, liu2024real, zhao2025mm}. However, as illustrated in Figure~\ref{fig:teaser}, deploying HPE in these environments introduces two critical constraints. First, \textbf{privacy sensitivity} renders vision-based sensors unusable, as residents are unwilling to accept continuous video monitoring in intimate spaces, regardless of downstream data security protocols. Second, \textbf{resource scarcity} precludes the use of computationally intensive inference models; residential edge devices typically lack GPU acceleration and have limited memory, yet the system must deliver estimates within hard deadlines to ensure timely interventions. Thus, the core research challenge lies in achieving a robust balance between privacy and utility: identifying a sensing modality that inherently preserves \textbf{privacy}, and architecting a processing pipeline that guarantees \textbf{schedulable}, bounded execution on low-power architectures.


Among emerging technologies, Millimeter-wave (mmWave) radar stands out as a natural candidate that satisfies both requirements in principle. Its RF-only sensing captures no identifiable visual content, eliminating the privacy concerns inherent to cameras \cite{zhao2018rf,guhr2020privacy,alshehri2022exploring}. Crucially, mmWave outputs a structured \textbf{Range-Angle-Doppler (RAD)} tensor that naturally mitigates resource scarcity and provides a foundation for execution predictability \cite{li2024sbrf,deng2023midas++}. Specifically, human reflections only occupy a distinct fraction of this tensor, inherently constrained by physical body dimensions (\emph{spatial compactness}) and non-zero velocities (\emph{motion distinctiveness}). Exploiting these properties eliminates redundant computation over static background clutter and bounds the processing workload to physical constants. This structural conversion to mathematically deterministic execution directly enables schedulable HPE on devices with limited resources.

Despite this theoretical potential, existing mmWave pipelines fail to translate structural advantages into deployable mobile HPE systems. Current approaches generally fall into two traps:

\textbf{1) Uniform dense processing:} Systems applying complex neural networks across the entire RAD tensor treat all bins as uniformly meaningful \cite{kini2025millimamba,wu2024mmhpe,zhao2025mm,sang2026ifdnet}, wasting computational cycles on the 75–90\% of empty bins \cite{zheng2025person,zheng2026learn}. Because tensor volume scales with the scene, this forces unbounded workload scaling in multi-person scenarios rather than isolating bounded per-instance subproblems. Consequently, execution time becomes a volatile function of scene complexity, causing unpredictable latency that violates the hard deadline guarantees required by high-response mobile schedulers.

\textbf{2) Blind lossy compression:} Systems uniformly subsampling the RAD tensor to reduce computational load treat all values as interchangeable \cite{yataka2024retr,fan2024diffusion,liu2024real}. Applied without regard to mmWave's inherent structure, they discard vital, physics-grounded Doppler signatures along with noise, severely degrading accuracy. Furthermore, these static reductions leave the system with a \textbf{lack of schedulability}, offering no runtime mechanism to adapt accuracy to available timing budgets, ultimately defeating the deployment's purpose.

Taken together, these two oversights produce three interconnected system-level failures. First, \textbf{unpredictable latency} arises because execution time remains coupled to scene complexity rather than to analytically bounded worksets. Second, \textbf{unbounded workload scaling} emerges because each additional person expands processing over the global scene representation instead of introducing a bounded per-instance subproblem. Third, \textbf{lack of schedulability} persists because most existing systems provide no principled mechanism for adapting accuracy to the timing budget available at runtime. These three failures define the requirements that the remainder of the paper addresses: predictable operator cost, stable scaling with target count, and deadline-aware runtime control.

To resolve these oversights, we present Physics-bounded Real-world Inference for mmWave Skeleton Modeling (PRISM), a framework that exploits a unified physical insight to achieve efficiency and analyzability simultaneously. PRISM is built around three coordinated components:
1) \textbf{Physics-Bounded Integral Processing (PBIP)} restricts feature extraction to physically occupied spatial and Doppler regions and replaces sliding windows of variable cost with integral queries that execute in constant time. This eliminates wasted CPU cycles, thereby transforming processing cost from a scene-dependent runtime variable into a closed-form function of measurable physical parameters.
2) \textbf{Physics-Adaptive Instance Proposal (PAIP)} addresses unbounded scaling by leveraging preserved Doppler signatures to decompose scenes into bounded subproblems for each person, ensuring workload scales strictly linearly with target count.
3) \textbf{Deadline-Aware Operation Profiles (DAOP)} converts the analytical bounds of PBIP and PAIP into a schedulable runtime interface. DAOP dynamically selects the most accurate operating profile that satisfies the deadline at each frame, based on worst-case execution times (WCET) verified offline.

Our contributions are summarized as follows:
\begin{itemize}
    \item We identify a schedulability gap in mmWave HPE for resource-constrained mobile edge deployment. Existing pipelines either process the full RAD tensor with scene-dependent latency or apply indiscriminate compression that removes pose-relevant motion structure. We distill three requirements that follow from this gap: timing predictability, linear workload scaling with person count, and deadline-aware runtime control.
    \item We present PRISM, a physics-bounded framework that converts spatial compactness and motion distinctiveness into analytically bounded computation. PBIP replaces variable-cost aggregation with constant-time integral queries, PAIP decomposes multi-person scenes into bounded per-instance subproblems, and DAOP selects among offline-verified operating profiles, returning the most accurate feasible estimate under each frame deadline.
    \item We evaluate PRISM on a Raspberry Pi~5 across four public datasets that span radar front-ends, single- and multi-person occupancy, and complementary pose supervision. Under isolated single-threaded execution, PRISM meets per-frame deadlines on the evaluated traces while remaining the most accurate among deadline-feasible configurations, supporting schedulable, privacy-preserving sensing on mobile edge hardware.
\end{itemize}

\section{Related Work}
\label{sec:related}

\textbf{mmWave HPE with Heavy Learned Front-Ends.}
Recent mmWave HPE systems increasingly adopt deep front-ends that process full or lightly pruned RAD tensors to improve reconstruction quality \cite{chang2020spatial,sengupta2022mmpose, mei2024mmspyvr, palipana2021pantomime,yu2026dynamic}. These systems often report competitive accuracy, but their dominant cost is still dense feature extraction over large tensors, and this cost changes with scene density and model depth \cite{zheng2025person,zheng2025differentiable,niu2015survey,peng2026enabling}. As a result, they typically lack analyzable worst-case execution time (WCET) bounds under data-dependent inference cost, and their workload grows unboundedly as active targets increase, so accuracy is attained only when timing constraints are relaxed.

\textbf{Classical and Physics-Aware Radar Processing.}
Another line of work relies on hand-crafted signal processing, including thresholding, beam-space processing, and geometry-driven feature engineering \cite{kini2025millimamba,zhu2024probradarm3f}. Compared with heavy learned front-ends, these approaches are often lighter and more interpretable, and some systems explicitly exploit radar-domain structure \cite{choi2025mvdoppler,zheng2026learn, zheng2026doppler}. Their common limitation is a constrained trade-off surface: aggressive compression may meet timing bounds but discards pose-relevant information; alternatively, preserving information without workset restriction leaves multi-person scaling unresolved. Formal end-to-end WCET guarantees remain uncommon in this category.

\textcolor{black}{\textbf{Why Heatmap-Level Reasoning Rather Than On-Chip Point Clouds.}
A natural question is why this work targets the dense Range-Angle-Doppler tensor at all, given that commodity FMCW radar systems-on-chip (e.g., the TI IWR/AWR18xx family) already perform range-Doppler FFT, CFAR detection, and angle-of-arrival estimation on integrated on-chip DSP hardware, exporting a compact point cloud at negligible host-CPU cost and low transfer bandwidth. We do not dispute this hardware advantage: on-chip CFAR is substantially faster than any host-side alternative, including PRISM's own bounded operators, and a point-cloud front-end will always win a narrow host-latency comparison against heatmap-level processing. The reason PRISM nonetheless operates on the pre-CFAR tensor is that the on-chip detection threshold is calibrated for generic object detection rather than pose estimation, and it necessarily discards sub-threshold reflections, precisely the low radar-cross-section, small-amplitude returns from limbs and extremities that carry the Doppler-resolved motion structure identified in O2. This information loss is empirically visible in Section~\ref{sec:eval_e2e}: mmDiff, which consumes a CFAR-derived point cloud, attains the weakest MPJPE and PCK@100mm among all evaluated methods despite its point-cloud front-end incurring negligible generation cost relative to its own inference stage. PRISM's contribution is therefore not a claim of winning a timing race against on-chip point-cloud generation, a race that heatmap-level processing structurally cannot win and does not attempt to; it is the narrower and directly supported claim that the richer, higher-bandwidth heatmap representation, previously considered incompatible with hard real-time constraints on a GPU-less host, can be rendered schedulable through physics-bounded computation without discarding the sub-threshold structure that on-chip point-cloud generation sacrifices at the sensor.}

\textbf{Real-Time Adaptive Inference Systems.}
Real-time AI research has proposed adaptation paradigms such as early-exit inference, anytime computation, and dynamic compute allocation under deadlines \cite{kong2024survey}. These frameworks provide important control ideas for timeliness and runtime quality trade-offs \cite{zheng2026learn}. For mmWave HPE, however, adaptation policies are usually model-centric rather than sensing-physics-centric; without explicit RAD workset constraints, their timing models are not naturally coupled to occupancy and Doppler bounds. Recent work on DNN inference WCET~\cite{salehzadeh2024wearable} has bounded costs for specific layer classes, but these bounds cover inference only, not the sensor-to-output latency chain, and do not structurally isolate scene complexity. The WCET community has developed both static analysis and measurement-based probabilistic timing analysis (MBPTA) with extreme-value theory (EVT) fitting~\cite{cucu2012measurement}; however, existing MBPTA applications do not structurally exploit sensing-physics loop bounds to decompose scene-level complexity. Consequently, layer-level timing bounds may hold while sensor-to-output latency and multi-person scaling remain unaddressed.

\textbf{Profile-Based Runtime Control for Edge AI.}
Profile-based controllers and runtime quality-of-service mechanisms expose discrete operating points for latency-accuracy trade-offs \cite{sun2024pbphs}. This pattern is compatible with real-time deployment because profile costs can be bounded offline and selected online with low overhead \cite{sheraz2020artificial}. Existing profile systems, however, are mostly built for generic vision or neural stacks; they rarely encode mmWave-specific physical bounds, namely occupancy-constrained spatial support and velocity-constrained motion support. Without physics-bounded operators underneath, profile selection cannot keep WCET data-independent, even when accuracy--latency trade-offs are exposed, and linear scaling with person count is not structurally enforced.

\textbf{Positioning of This Work.}
None of the above categories provides an integrated chain from sensing physics to analyzable scheduling behavior that jointly supports timing-predictable operators, bounded per-person workloads, and deadline-feasible profile control. Section~\ref{sec:formulation} formalizes these three properties as system objectives, respectively denoted R1 (timing predictability), R2 (accuracy maximization under timing constraints), and R3 (scalable stability); PRISM occupies the gap through the coordinated design of PBIP, PAIP, and DAOP.

\section{mmWave HPE Foundations and Observations}
\label{sec:background}

Section~\ref{sec:intro} identified the spatial compactness and motion distinctiveness of mmWave. If these properties are exploited, mmWave should enable schedulable HPE on platforms with limited resources. This section provides the technical foundations and observations for these claims. 

\subsection{RAD Tensor Formation}
\label{sec:primer}

A Frequency-Modulated Continuous-Wave (FMCW) mmWave radar transmits a sequence of chirps whose frequency increases linearly over time \cite{iovescu2020fundamentals}. Reflected signals are captured by an antenna array and processed through a cascaded Fast Fourier Transform (FFT) pipeline that extracts three physical dimensions \cite{richards2005fundamentals}:

\begin{itemize}
    \item \textbf{Range ($R$):} Mixing the received signal with the transmitted chirp produces an Intermediate Frequency (IF) signal whose frequency encodes round-trip delay. A Range FFT converts this to distance.
    \item \textbf{Doppler ($D$):} Radial motion induces phase shifts between consecutive chirps. A Doppler FFT across the chirp sequence yields radial velocity.
    \item \textbf{Angle ($A$):} Phase differences across spatially distributed antennas encode direction. An Angle FFT translates these differences into azimuth and elevation.
\end{itemize}

\begin{figure}[t]
  \centering
  \includegraphics[width=0.9\linewidth]{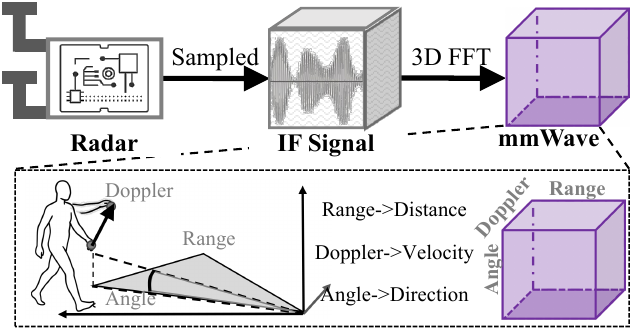}
  \caption{mmWave radar workflow and the physical characteristics of mmWave in each dimension.}
  \label{fig:mmwave}
\end{figure}

The output is a structured tensor $\mathcal{T} \in \mathbb{R}^{R \times A \times D}$, where each index triple $(r, a, d)$ corresponds to a specific spatial volume moving at a specific velocity, as illustrated in Figure~\ref{fig:mmwave}. This deterministic mapping between data coordinates and physical properties is the foundation for physics-bounded processing: computation can be confined to regions where valid human reflections exist.

\subsection{HPE Deployment Environment}
\label{sec:system_model}

The target HPE platform is an embedded CPU or low-power SoC without GPU acceleration, imposing an upper bound on computational throughput, $F_{\text{platform}}$. Radar frames arrive at period $\tau$, and pose estimates should be completed before a hard deadline $T_d \le \tau$. Section~\ref{sec:formulation} formalizes the resulting task model and timing constraints.

\subsection{Empirical Observations}
\label{sec:observations}

To quantify the physical bounds, we profile RAD tensors acquired during human motion across multiple indoor environments, yielding four observations and one auxiliary property.

\textbf{O1 (Spatial Compactness):}
The human body subtends a bounded solid angle at typical indoor radar ranges: its physical cross-section in the range-angle plane is constrained by body width ($\lesssim$1~m) and the radar's angular resolution \cite{rahman2024mmvr}, so valid reflections occupy only a compact fraction $\rho_s \ll 1$ of the full $(R \times A)$ plane. The remaining bins contain static structures or noise with no pose-relevant content. Processing can therefore be restricted to the occupied region $\Omega_s$ without information loss. The specific $\rho_s$ fractions realized across the four public evaluation datasets are reported in Section~\ref{sec:eval_bounds}, confirming this physics prediction under diverse radar configurations.

\textbf{O2 (Motion Distinctiveness):}
Stationary objects (walls, furniture, and floor) emit energy at or near zero Doppler shift \cite{choi2025mvdoppler}. Human limb and torso motion is biomechanically bounded: ambulatory and gesture velocities span a predictable non-zero range, mapping to a kinematically determined fraction $\rho_d < 1$ of the Doppler axis whose exact extent depends on the radar's velocity resolution. This separation is kinematic, not geometric: regardless of room layout, static clutter concentrates at zero Doppler while human motion occupies a distinct non-zero band. The measured $\rho_d$ fractions across the four public evaluation datasets are reported in Section~\ref{sec:eval_bounds}. 
%

\textbf{O3 (Timing Bottleneck):}
Standard feature extraction uses sliding windows over $\mathcal{T}$. The cost of each query scales as $(2w{+}1)^2$ with window half-width $w$. Profiling confirms super-linear growth: doubling $w$ from 2 to 4 increases the cost for each frame by $\approx 3.5\times$ on the target platform \cite{yataka2024retr}. Multi-person scenes compound this cost when window statistics are recomputed for each instance. Eliminating the $w$-dependence is necessary to bound worst-case latency.

\textbf{O4 (Stability Across Scenes):}
We measure overlap between Doppler masks calibrated independently across different rooms with distinct layouts and clutter distributions. The zero-Doppler static band remains consistent: a mask calibrated in any one room suppresses $>$97\% of static energy in the remaining rooms without retuning, as quantified in Section~\ref{app:cross_scene}. This stability arises because Doppler encodes velocity, a property of motion rather than scene geometry. A single physics-derived threshold therefore transfers across deployment sites.

\textbf{AP (Accuracy–Scope Monotonicity):} 
Coarser spatial or Doppler bounds yield lower joint localization accuracy. This monotonic relationship enables principled profile ordering: higher-cost profiles process larger scopes and achieve higher accuracy, supporting deadline-aware trade-offs.

These observations establish that human reflections reside in a schedulable, physics-bounded subset of the RAD tensor. The mapping to system design is:
\begin{itemize}
    \item \textbf{O1+O2+O3→PBIP:} Spatial and Doppler bounds define the valid processing region; eliminating window-size dependence within this region yields constant-time queries.
    \item \textbf{O2+O4→PAIP:} Motion distinctiveness enables target–background separation; stability across scenes ensures the separation criterion transfers without retuning.
    \item \textbf{AP→DAOP:} Monotonic accuracy–scope relationship enables ordered profile selection under deadline constraints.
\end{itemize}

\section{Problem Formulation and System Objectives}
\label{sec:formulation}

Section~\ref{sec:background} established that human reflections occupy bounded supports (O1, O2), that window-based aggregation is the timing bottleneck (O3), that the Doppler separation criterion generalizes across scenes (O4), and that accuracy degrades monotonically with processing scope (AP). 
This section translates these empirical bounds into system objectives that directly target the three flaws: unpredictable latency, unbounded workload scaling, and lack of schedulability.

\subsection{Task Model}

We consider a continuous sensing system that processes a stream of RAD tensors $\{\mathcal{T}_k\}$ arriving at period $\tau$. For each frame $k$, the system should produce pose estimates for up to $M$ human targets before a hard deadline $T_d \le \tau$. The computational workload depends on tensor dimensions $(R, A, D)$, target count $M$, and the processing scope parameters $(\rho_s, \rho_d)$ defined in O1 and O2.
To enable runtime adaptation, the system exposes a discrete set of operating profiles $\mathcal{P}$. Each profile $p \in \mathcal{P}$ specifies a configuration $(\rho_s^{(p)}, \rho_d^{(p)}, M_{\max}^{(p)})$ that determines both computational cost and expected accuracy.

\subsection{System Objectives}


\noindent\textbf{R1 (Timing Predictability).} \\
The system should provide, for every profile, a verifiable worst-case execution time (WCET) that upper-bounds that profile's own execution regardless of input content, under the deployment assumptions of single-threaded execution with fixed CPU frequency and process isolation:
\begin{equation}\label{eq:R1}
\small
    T(\mathcal{T}_k, p) \le T_{\text{WCET}}^{(p)}, \quad \forall\, \mathcal{T}_k, \;\forall\, p \in \mathcal{P}.
\end{equation}
R1 requires only that $T_{\text{WCET}}^{(p)}$ be a valid, input-independent bound for each profile in isolation; it does not require every profile to satisfy an arbitrary deadline $T_d$, since low-cost and high-cost profiles are designed to serve different timing budgets. Whether a specific profile additionally meets the deadline in force, i.e., $T_{\text{WCET}}^{(p)} \le T_d$, is the feasibility test applied by DAOP's selection rule in R2 below, evaluated only over profiles under consideration for selection. The bound $T_{\text{WCET}}^{(p)}$ should be a closed-form function of physics-derived parameters $(\rho_s^{(p)}, \rho_d^{(p)}, M_{\max}^{(p)}, N_q)$, where $N_q$ is the fixed number of PBIP box queries issued per retained instance, together with platform constants calibrated via worst-case micro-benchmarks. This construction, consistent with the Measurement-Based Probabilistic Timing Analysis (MBPTA) paradigm~\cite{cucu2012measurement}, encodes the physics-derived loop counts directly rather than relying on distributional extrapolation from runtime execution traces. Under unmanaged co-runner contention, the effective margin may need to be increased, as examined in Section~\ref{app:limitations}.

\noindent\textbf{R2 (Accuracy Maximization under Timing Constraints).} \\
Given R1\&R3 as hard constraints, the system should maximize pose accuracy by selecting the highest-quality feasible profile:
\begin{equation}\label{eq:R2}
\small
    p^* = \arg\max_{p \in \mathcal{P}} \;\text{Accuracy}(p) \quad \text{s.t.} \quad T_{\text{WCET}}^{(p)} \le T_d.
\end{equation}
This formulation requires that the profile set $\mathcal{P}$ be ordered by both cost and accuracy (enabled by AP), and that selection be performed at runtime with negligible overhead. The result is a schedulable interface: the system can trade accuracy for timing margin when deadlines tighten or resources shrink.

\noindent\textbf{R3 (Scalable Stability).}\\
Execution time should grow linearly with target count rather than quadratically, and should not depend on scene-specific clutter:
\begin{equation}\label{eq:R3}
\small
    T_{\text{WCET}}(M) \le C_0 + M \cdot C_1, \quad M \le M_{\max}^{(p)},
\end{equation}
where $C_0$ is a bounded constant for scene-independent preprocessing and $C_1$ is the per-instance cost upper bound. This formulation ensures that adding one person increases cost by at most $C_1$, preventing the super-linear growth observed in existing pipelines.

\subsection{Objective-to-Design Mapping}

Three objectives map to architectural components as:

\textbf{(1) R1 → PBIP:} Timing predictability requires eliminating the window-size dependence identified in O3. PBIP achieves this by replacing $(2w{+}1)^2$-cost aggregation with $O(1)$ integral queries over the bounded supports from O1 and O2.

\textbf{(2) R3 → PAIP:} Scalable stability requires isolating the workload for each person. PAIP leverages the motion distinctiveness from O2 and the stability across scenes from O4 to decompose the scene into bounded local subproblems for each instance.

\textbf{(3) R2 → DAOP:} Accuracy maximization under deadline constraints requires runtime profile selection. DAOP exploits the monotonic accuracy–scope relationship from AP to order profiles and selects the best feasible option in $O(1)$ time.

\begin{figure*}[t]
  \centering
  \includegraphics[width=\linewidth]{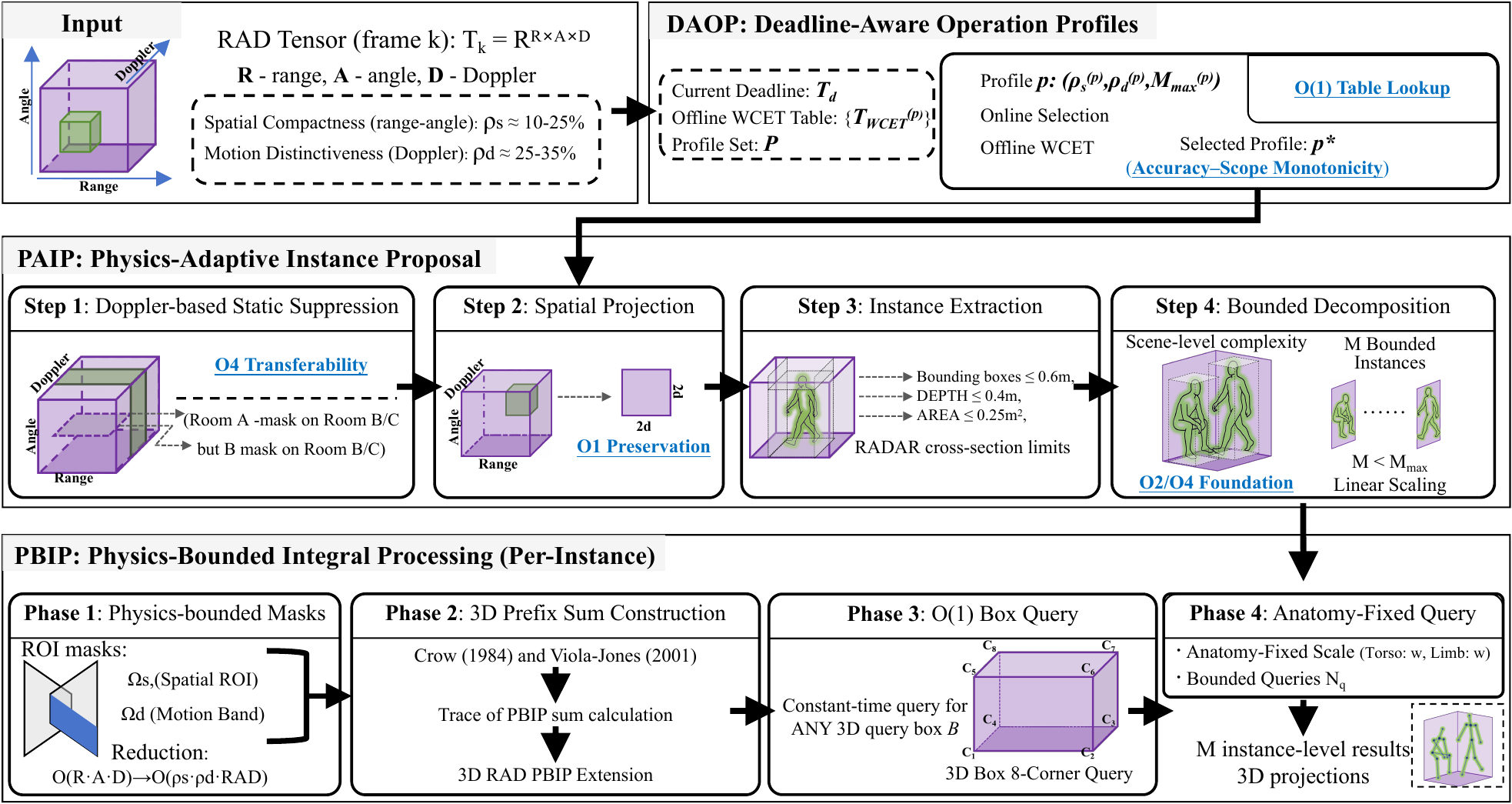}
  \caption{Overview of the PRISM pipeline. Each radar frame is a RAD tensor whose physics-measurable compactness fractions bound the valid processing region. \textbf{DAOP selects} the highest-accuracy profile whose offline-verified WCET satisfies the current deadline via an $O(1)$ table lookup over the monotone profile set. \textbf{PAIP}: the selected profile's parameters drive a four-step instance proposal. \textbf{PBIP}: for each instance sub-tensor, physics-bounded masks restrict the support, a 3D prefix-sum is built once and reused, and $N_q$ anatomy-fixed $O(1)$ box queries produce a fixed-length feature vector fed to the MLP regressor.}
  
  \label{fig:pipeline_overview}
\end{figure*}

Figure~\ref{fig:pipeline_overview} illustrates the end-to-end information flow through these three components.

\section{Physics-Bounded System Architecture}
\label{sec:design}


\subsection{Deadline-Aware Operation Profiles (DAOP)}
\label{sec:design_profiles}

\textbf{Motivation for Online Profile Selection.}
A natural question is whether a single profile, selected offline for the worst-case operating condition, would suffice. Two complementary arguments show that it does not. 
\textbf{First}, the per-frame computational budget available on a shared mobile edge platform is not constant. Co-resident system tasks (e.g., sensor drivers, network daemons) cause transient CPU contention that can temporarily reduce available slack, and the upper-level scheduler may impose different per-task time budgets under varying system load. Frame-to-frame variation in the detected person count $M$ changes the realized cost of PAIP and PBIP under the already selected $M_{\max}$, but it does not revise the profile decision.
\textbf{Second}, fixing a conservative profile calibrated for the tightest deadline permanently sacrifices accuracy when a looser budget is available.

The combined effect is that a fixed single-profile system simultaneously risks deadline violations under tighter timing budgets and delivers sub-optimal accuracy under relaxed ones. DAOP resolves both issues through online profile selection: it selects the highest-accuracy feasible profile at each frame, given the current deadline, without requiring any runtime execution-time measurement.

\textbf{Role and Physical Basis.}
DAOP is the runtime control layer: before each frame is processed, it selects an operating profile $p^*$ that configures the scope parameters $(\rho_s, \rho_d, M_{\max})$ consumed by the downstream PAIP and PBIP stages (\S\ref{sec:design_paip}--\S\ref{sec:design_pbip}). The Auxiliary Property (AP) confirms that pose accuracy degrades monotonically with reduced scope, enabling principled profile ordering: given a deadline, DAOP always selects the highest-accuracy configuration whose worst-case cost is guaranteed not to exceed $T_d$. The accuracy ordering $\text{Accuracy}(p_1)\le\cdots\le\text{Accuracy}(p_K)$ is established per deployment during the same offline characterization that derives $(c_1,c_2,c_3)$; stability across scenes in this ordering is empirically supported by the MPJPE variation of $\pm$3.8~mm across scenes in the validation split in Section~\ref{app:cross_scene}, confirming that profile ranking does not invert under realistic environment variation.

DAOP adopts a discrete profile variant: rather than continuous adjustment, it selects among pre-characterized configurations, each with an offline-verified WCET bound. This discretization avoids runtime profiling overhead and guarantees deadline satisfaction by construction.
We define ordered profiles $\mathcal{P} = \{p_1, p_2, \ldots, p_K\}$, where each profile $p$ specifies $(\rho_s^{(p)}, \rho_d^{(p)}, M_{\max}^{(p)})$. Profiles are monotone by construction: higher-indexed profiles allocate larger worksets, have non-decreasing WCET, and achieve non-decreasing expected accuracy. The granularity of $\mathcal{P}$ is a deployment-time design choice: finer spacing reduces the worst-case accuracy loss per profile downgrade at the cost of a larger offline characterization effort. Concretely, the five profiles reported in this paper are obtained by a grid search over $(\rho_s,\rho_d,M_{\max})$ that traces the offline WCET--accuracy Pareto frontier, from which five well-separated, representative operating points are retained as $\mathcal{P}$.

\textbf{Offline WCET Construction.}
PRISM adopts the Measurement-Based Probabilistic Timing Analysis (MBPTA) paradigm~\cite{cucu2012measurement}: analytic loop bounds are supplied externally from physics ($\rho_s$, $\rho_d$, $N_q$, $M_{\max}$), and the remaining micro-architectural variance is captured by platform constants $(c_1, c_2, c_3)$ derived from worst-case micro-benchmarks. The calibration procedure and margin policy are detailed in the Experimental Setup (\S\ref{sec:eval_setup}). For profile $p$, $T_{\text{WCET}}^{(p)} \le $:
\begin{equation}\label{eq:WCET}
\small
  \frac{c_1 R A D + c_2\rho_s^{(p)}\rho_d^{(p)} R A D + c_3 M_{\max}^{(p)}\!\left(A_{\text{person}}D + N_q\right)}{F_{\text{platform}}} + T_{\text{sw}},
\end{equation}
where $T_{\text{sw}}$ is profile switching overhead, $A_{\text{person}}$ denotes the bounded range-angle support allocated to one person, and $N_q$ is the fixed number of PBIP queries issued for each retained instance. The three numerator terms cover the \emph{complete} host-side sensor-to-output pipeline: $c_1 R A D$ is PAIP's scene-level scan and Doppler projection cost over the incoming RAD tensor; $c_2\rho_s^{(p)}\rho_d^{(p)} R A D$ is PBIP's bounded-support summed-volume construction cost (\S\ref{sec:design_pbip}); and $c_3 M_{\max}^{(p)}(A_{\text{person}}D + N_q)$ is the upper bound for each instance, combining PAIP's local tensor extraction with PBIP's fixed-count query and regression stage. The front-end tensor is produced upstream by the sensor pipeline or dataset front-end, so this formula covers the host-side computation actually controlled by PRISM. Because all cost terms are deterministic functions of physics parameters, the resulting WCET table does not rely on online estimation.

\textbf{Online Selection Rule.}
At frame $k$, given deadline $T_d$, DAOP instantiates the selection rule of Eq.~\eqref{eq:R2}, returning the highest-accuracy feasible profile $p_k^*$.
Since $|\mathcal{P}|$ is fixed and small, selection is an $O(1)$ table lookup. If available slack decreases, DAOP switches to a lower profile rather than violating deadlines. If even the lightest profile $p_1$ exceeds $T_d$, the frame is declared unschedulable and dropped; persistent infeasibility signals a capacity planning failure rather than a runtime failure of PRISM. Because PAIP and PBIP reconstruct data structures for each frame, switching profiles incurs no state-flushing cost; the overhead $T_{\text{sw}}$ is bounded and included in Eq.~\eqref{eq:WCET}.

\subsection{Physics-Adaptive Instance Proposal (PAIP)}
\label{sec:design_paip}

\textbf{Decision Order and Dependency Structure.}
To prevent any ambiguity, we state the runtime execution order explicitly. 
\textbf{DAOP is invoked \emph{first}}, before any frame data are processed, and establishes the operating profile $p^*$ together with its associated budget cap $M_{\max}^{(p^*)}$. 
\textbf{PAIP executes \emph{second}} using the scope parameters $(\rho_s^{(p^*)}, \rho_d^{(p^*)}, M_{\max}^{(p^*)})$ provided by DAOP, and detects the actual number of proposals $M$ from the current frame's radar data. If $M > M_{\max}^{(p^*)}$, lower-energy proposals are truncated in Step~4. 
\textbf{PBIP then executes \emph{third}}, processing each of the up to $M_{\max}^{(p^*)}$ retained instances independently. 
The information flow is strictly unidirectional—DAOP $\rightarrow$ PAIP $\rightarrow$ PBIP—with no feedback from the detected person count to the profile selection decision, which has already been committed before frame processing begins. $M_{\max}^{(p^*)}$ is therefore a budget ceiling established by the profile, not a prediction of the actual scene occupancy.

Given the profile $p^*$ selected by DAOP, PAIP converts the global RAD tensor into a bounded set of local tensors, one for each instance, each processed independently by PBIP (\S\ref{sec:design_pbip}). R3 requires that execution time scale linearly with target count rather than with scene volume. PAIP achieves this by leveraging O2 and O4: O2 establishes that static objects concentrate near zero Doppler while human motion occupies a bounded non-zero Doppler band; O4 confirms that this Doppler-based separation generalizes across environments without recalibration for each scene.

The proposal-then-verify paradigm is common in object detection \cite{pelhan2024dave}. PAIP adapts this paradigm to radar by exploiting the Doppler axis for motion-based foreground extraction. The Doppler gate provides a criterion grounded in physics and independent of the scene for isolating human targets, avoiding learned or heuristic background models.

PAIP comprises four steps, each with an explicit physical rationale.

\textit{Step 1: Doppler-based static suppression.}
O2 establishes that stationary objects produce energy at or near zero Doppler. To isolate human motion:
\begin{equation}
\small
    \mathbf{R}_{\text{active}}[r,a,d] = \mathcal{T}_k[r,a,d]   \big[|v(d)| > v_{\text{static}}\big],
\end{equation}
where $v(d)$ maps Doppler bin $d$ to radial velocity and $v_{\text{static}}$ is a threshold derived from the minimum human motion velocity. O4 confirms that this threshold transfers across rooms without retuning.


\textit{Step 2: Spatial projection.}
The filtered tensor is projected onto the range-angle plane to consolidate motion evidence:
\begin{equation}
\small
    P_{\text{active}}[r,a] = \sum_{d\in\mathcal{D}_{\text{dyn}}} \left|\mathbf{R}_{\text{active}}[r,a,d]\right|^2.
\end{equation}
Here $\mathcal{D}_{\text{dyn}} = \{d:\, |v(d)| > v_{\text{static}}\}$ is the dynamic Doppler-bin set retained after Step~1. This projection reduces dimensionality while preserving spatial localization of moving targets.

\textit{Step 3: Anthropometric-constrained proposal extraction.}
Connected components on $P_{\text{active}}$ produce candidate regions. Each candidate is evaluated against a two-sided physics-derived envelope. A \emph{lower bound} $A_{\min}$ on component area rejects isolated noise and micro-reflections too small to correspond to any human body part. An \emph{upper bound} $L_r \times L_a$ rejects regions whose range-angle footprint exceeds the worst-case adult body envelope derived from 97.5th-percentile anthropometric data~\cite{sun2024pbphs}; any single-body radar return must fit within this envelope regardless of posture, so regions beyond the threshold represent merged multi-person clusters or static objects that leaked through the Doppler gate. An additional ceiling $A_{\max}$ on total active-bin count provides a second rejection criterion for diffuse clutter.

\textit{Step 4: Bounded instance decomposition.}
Let validated regions be $\{\mathcal{B}_m\}_{m=1}^{M}$. If $M > M_{\max}^{(p^*)}$, proposals are ranked by projected energy $\sum P_{\text{active}}[r,a]$ within each component, and only the top-$M_{\max}^{(p^*)}$ are retained; lower-energy proposals are dropped to preserve the WCET guarantee. For each retained $\mathcal{B}_m$, PAIP extracts a local sub-tensor $\mathcal{T}_{k,m}$ and passes it to PBIP (\S\ref{sec:design_pbip}) for independent processing. This converts scene-level variability into a bounded set of instance-level tasks.


When truncation occurs in higher-occupancy settings, the energy-based ranking policy preferentially retains persons with stronger radar reflections, which generally correspond to persons closer to the sensor or exhibiting more prominent motion. Persons at the margin of the Doppler gate (near-stationary targets) or at longer range would tend to have weaker projected energy and be dropped first. In tracking-oriented deployments where continuity across frames is required, temporal propagation of proposals from preceding active frames can bridge brief periods in which a low-energy person is truncated, though this extension is outside the scope of the per-frame schedulability model described here.

\textbf{Complexity Guarantee.}
The Doppler-gating and projection steps traverse the incoming RAD tensor once for each frame, whereas only the local extraction stage repeats across retained proposals. PAIP complexity therefore separates into one scene-level term and one instance-level term:
\begin{equation}
\small
  C_{\text{PAIP}} = O(R A D) + O\!\left(M\cdot A_{\text{person}}\cdot D\right), \quad M\le M_{\max}^{(p^*)},
\end{equation}
where $A_{\text{person}}$ is the bounded spatial support for each person. The first term is the scene scan performed once for each frame that corresponds to the $c_1 R A D$ term in Eq.~\eqref{eq:WCET}, and the second upper-bounds the repeated local extraction cost. Because only the second term scales with $M$, execution time still increases linearly with target count, satisfying the stability condition required by R3.

\begin{figure}[t]
  \centering
  \includegraphics[width=\columnwidth]{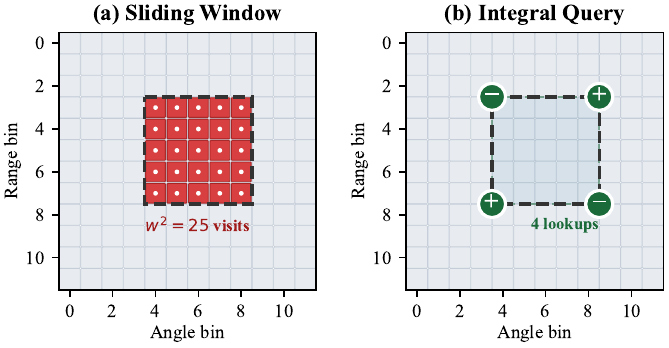}
  \caption{Comparison of query operators on a range-angle tensor slice.
  \textbf{(a)}~Sliding window: every element inside the $w{\times}w$ region
  (red cells) must be visited, costing $O(w^2)$ per query, and cost rises with
  window size and is therefore \emph{data-dependent}.
  \textbf{(b)}~PBIP integral query: after a prefix-sum construction performed once
  over the physics-bounded support, any region sum reduces to four corner
  lookups (circles, $+$/$-$ denote inclusion--exclusion signs), costing $O(1)$
  regardless of window size. This $w$-independence converts execution time for each frame
  into a closed form, analytically bounded expression.}
  \label{fig:sliding_vs_integral}
\end{figure}

\subsection{Physics-Bounded Integral Processing (PBIP)}
\label{sec:design_pbip}
 

PBIP is the bounded-support feature extraction stage that follows PAIP. Its summed-volume table is constructed once per frame over the masked tensor, after which each retained instance issues a fixed set of constant-time queries. It addresses the timing bottleneck identified in O3: window-based local aggregation scales as $(2w{+}1)^2$ with window half-width $w$, causing execution time to vary with query geometry (Figure~\ref{fig:sliding_vs_integral}(a)). PBIP eliminates this dependence by replacing per-query enumeration with constant-time region lookups over a precomputed integral structure (Figure~\ref{fig:sliding_vs_integral}(b)). The physical grounding comes from O1 and O2: because human reflections occupy only a bounded fraction of the spatial plane ($\rho_s$) and Doppler axis ($\rho_d$), the integral structure need only be built over this bounded support rather than the full tensor.


\textbf{Physics-Driven Adaptations.}
A direct summed-volume construction over the full RAD tensor does not resolve the timing problem: preprocessing cost remains $O(R A D)$, query region extents depend on data content rather than physical constants, and WCET cannot be derived as a closed form expression. Because $\rho_s\rho_d \ll 1$ in typical mmWave deployments (Table~\ref{tab:bounds}), restricting construction to the physics-bounded support reduces the element count by $1/(\rho_s\rho_d)$, typically by an order of magnitude. This makes construction cost a closed form term that directly enables the WCET formula in Eq.~\eqref{eq:WCET}. PBIP introduces three coordinated adaptations that together convert the generic primitive into a timing certifiable operator:

\textit{A1 (Spatial ROI restriction).}
O1 establishes that valid human reflections occupy $\rho_s$ of the range-angle plane. PBIP restricts preprocessing to the spatial region $\Omega_s$ where human presence is physically plausible, reducing preprocessing cost by a factor proportional to $1/\rho_s$ relative to full-tensor processing.

\textit{A2 (Doppler band restriction).}
O2 establishes that human motion falls within a bounded Doppler band corresponding to $\rho_d$ of the Doppler axis. Static clutter near zero Doppler is excluded from the summed-volume structure. Preprocessing thus spans only $\rho_d D$ Doppler bins, further reducing cost by $1/\rho_d$.

\textit{A3 (Anthropometrically bounded query scales).}
Human body parts have bounded physical dimensions determined by anatomy. These dimensions map to fixed bin counts at the tensor's spatial resolution. Because query extents are constants derived from body geometry rather than data-dependent parameters, the number of queries per frame is bounded and each query is $O(1)$ after preprocessing. Concretely, $N_q = J \cdot |\mathcal{W}|$ per instance, where $J$ is the number of target joint types in the skeleton model and $|\mathcal{W}|$ is the number of anthropometrically-derived query scales per joint class; both are fixed at system design time. On the \textcolor{black}{17}-joint benchmark skeleton with \textcolor{black}{3} query scales per joint type, $N_q = \textcolor{black}{51}$ per instance, independent of scene complexity or person count. Each joint-type query box is anchored at a fixed offset from the instance's proposal centroid (Step~3 below), following the canonical joint layout of the evaluated dataset's skeleton definition, so anchor placement is a deterministic function of proposal geometry rather than a separately learned or per-frame estimated quantity. The three scales are chosen to span the range of apparent body-part sizes across the operational distance range, so that at least one query scale matches the target's extent under the current front-end resolution.

\textbf{Technical Definition.}
For frame $k$, let $\mathcal{T}_k \in \mathbb{R}^{R\times A\times D}$ and define a masked tensor
\begin{equation}\label{eq:masked}
\small
  \mathcal{X}_k[r,a,d] = \mathcal{T}_k[r,a,d] \cdot \mathbb{1}\left[(r,a)\in\Omega_s,\ d\in\Omega_d\right],
\end{equation}
where $\Omega_s$ and $\Omega_d$ are physics-bounded supports induced by $(\rho_s,\rho_d)$. $\Omega_s$ is an axis-aligned range-angle support whose extent is fixed by the profile's $\rho_s$ before frame processing, and $\Omega_d$ is a contiguous Doppler sub-band excluding the near-zero static band. The masked tensor $\mathcal{X}_k$ therefore visits exactly $\rho_s\rho_d R A D$ elements. Query boxes that straddle the ROI boundary are clamped to the support; zero padding preserves the correctness of inclusion-exclusion.

PBIP builds a summed-volume tensor $\mathcal{S}_k$ over $\mathcal{X}_k$:
\begin{equation}
\small
  \mathcal{S}_k[r,a,d] = \sum_{r'\le r} \sum_{a'\le a} \sum_{d'\le d} \mathcal{X}_k[r',a',d'],
\end{equation}
The sum over any axis-aligned query box $\mathcal{B}$ is then obtained by inclusion-exclusion with eight corner lookups:
\begin{equation}
\small
  \text{Sum}(\mathcal{B}) = \sum_{i=1}^{8} \sigma_i\,\mathcal{S}_k[c_i], \quad \sigma_i\in\{+1,-1\}.
\end{equation}
Per-query complexity is therefore $O(1)$, independent of window half-width $w$.

\textbf{Complexity Guarantee.}
Let the baseline sliding-window operator cost be $C_{\text{orig}} = O(\rho_s R A D \cdot (2w{+}1)^2)$. For a frame with $M$ retained instances, PBIP consists of one shared summed-volume construction over the bounded support and one fixed-count query stage per instance:
\begin{equation}
\small
  C_{\text{PBIP}} = O(\rho_s \rho_d\, R A D) + O(M\cdot N_q), \quad M\le M_{\max}^{(p^*)},
\end{equation}
where $N_q$ is the bounded number of queries per instance. The first term is paid once per frame, whereas the second is the repeated per-instance query cost. The critical property for R1 is that both terms are independent of window size $w$. Together with PAIP's $O(R A D) + O(M \cdot A_{\text{person}} \cdot D)$ decomposition cost, these terms directly constitute the numerator components of the WCET formula in Eq.~\eqref{eq:WCET}.

\textbf{Prefix-Sum Scope and Memory.}
The summed-volume tensor $\mathcal{S}_k$ is constructed \emph{once per frame} over the globally masked tensor $\mathcal{X}_k$ (Eq.~\ref{eq:masked}); all $M_{\max}N_q$ instance-level queries address this single shared table through constant-count corner lookups, with no per-instance allocation.

For peak-memory reasoning, the bound should include not only the shared prefix-sum table and the input tensor but also the PAIP intermediates that may coexist before query emission. A conservative upper bound is therefore:
$
  \mathcal{M}_{\text{peak}} = O\!\left(\rho_s \rho_d\, R A D\right)
  \;+\; O\!\left(R A D\right)
  \;+\; O\!\left(R A D\right)
  \;+\; O\!\left(R A\right)
  \;+\; O\!\left(M_{\max} \cdot N_q\right).
$
This expression counts the PAIP intermediates explicitly so that the bound remains valid even when the Doppler-gated tensor is materialized rather than overwritten in place. Because $\rho_s\rho_d \ll 1$ across our evaluation datasets, the global table still occupies only a small fraction of the dominant RAD-sized tensors; quantitative memory measurements are reported in the Experimental Setup (\S\ref{sec:eval_setup}).
%


\section{Evaluation}
\label{sec:eval}

This section provides quantitative closure on the three main claims that define PRISM's schedulability argument: operator-level timing properties (\S\ref{sec:eval_pbip}), multi-instance scaling behavior (\S\ref{sec:eval_paip}), and full-system profile control (\S\ref{sec:eval_daop}). 
Additional stress tests, including the PRISM-SWG ablation, EVT tail fitting, co-runner interference, a continuous occupancy case study, analyses of accuracy degradation, stability across scenes, and generalization across datasets are also reported.


\subsection{Experimental Setup}
\label{sec:eval_setup}

\textbf{Hardware Platform and Execution Modes.}
All experiments run on a Raspberry Pi~5 (BCM2712, quad-core Cortex-A76 at 2.4~GHz, 8~GB LPDDR4X, 1~MB L2 per core, 2~MB shared L3). CPU frequency scaling is \emph{disabled} (performance governor, fixed 2.4~GHz) to eliminate governor-induced latency variance; turbo is not available on this SoC. The platform runs a 64-bit Raspberry Pi OS. Unless otherwise stated, the inter-frame period is $\tau = 50$~ms (20~Hz) and the hard deadline is $T_d = 45$~ms. 

We evaluate PRISM under two complementary execution modes on this same platform. 
(1) In the \emph{dataset-replay mode}, the four public datasets below are streamed from local storage into the on-device pipeline; this mode isolates the algorithm from sensor-specific capture artifacts and provides a controlled, fully reproducible setting in which all baselines and PRISM are compared on identical input under the shared protocol (Sections~\ref{sec:eval_bounds}--\ref{sec:eval_e2e}). 
(2)In the \emph{live-capture mode}, the same Raspberry Pi~5 is connected to a TI AWR1843BOOST mmWave radar that streams frames in real time, and the on-device pipeline performs end-to-end acquisition, processing, and pose estimation under the same deadline; this mode validates that the schedulability and accuracy properties established in dataset-replay mode hold on a real sensor-to-output deployment (Section~\ref{sec:prototype}). The two modes share the identical PRISM implementation and trained model weights; they differ only in the source of the input frames.

\textbf{Datasets.}
Four public datasets are used, collectively designed to stress distinct dimensions of the PRISM design: WCET compliance under standard conditions, front-end generalization across signal representations, robustness across sensor generations, and accuracy fidelity under high-precision ground truth. Together, they establish that PRISM's physics-bounded guarantees are not artifacts of any single radar configuration. Each dataset is mapped onto a shared RAD lattice before processing, so that every method, including PRISM, consumes the same per-dataset representation~\cite{lee2023hupr,wang2024xrf55,ho2024rt,mueller2025radproposer}.

\textit{\textbf{HuPR}}~\cite{lee2023hupr}: The community benchmark for mmWave HPE, collected with a TI IWR1843 FMCW radar at 77~GHz producing dual-view heatmaps with angular resolution~$15^\circ$, 141K frames, 6~subjects, and a single-person setting. An explicit range--angle--Doppler representation is obtained by standard FMCW processing; pose supervision follows the released vision-aligned annotations~\cite{lee2023hupr}.

\textit{\textbf{XRF55}}~\cite{wang2024xrf55}: A multi-person indoor RF dataset whose mmWave front-end releases per-frame range--angle (RA) and range--Doppler (RD) maps. We reconstruct a RAD tensor from these two maps and use the same constructed tensor for every method; occupancy $M$ is taken from the synchronized Kinect skeleton stream released with the dataset, and pose supervision follows the authors' vision/depth annotations~\cite{wang2024xrf55}.

\textit{\textbf{\rtposeedit{RT-Pose}}}~\cite{ho2024rt}: \rtposeedit{A 4D radar tensor benchmark comprising 72K frames from 240 sequences across 40 indoor and outdoor scenes, with both single-person and multi-person labels. Pose supervision is the released RGB--LiDAR 3D skeleton~\cite{ho2024rt}. In our implementation, its released Cartesian 4D tensor is remapped into an RAD-aligned representation with an explicit Doppler axis before PRISM processing. We use its multi-person split as the primary workload-scaling and schedulability dataset because this remapped representation preserves the Doppler-resolved motion structure required by PAIP and PBIP while covering realistic scene diversity.} \textcolor{black}{Concretely, the released tensor is indexed by Cartesian spatial coordinates together with a velocity axis; we convert each Cartesian cell to polar range and azimuth through the standard geometric transform and resample onto the $R{\times}A$ lattice by bilinear interpolation, while the native velocity axis is retained directly as the Doppler dimension without resampling. This procedure preserves the full Doppler resolution of the source data, so the motion separability required by PAIP's Doppler gate is not degraded by the remapping. To prevent any representation-induced unfairness, the identical remapped tensor is supplied to every baseline that consumes a dense tensor, and the point-cloud baselines derive their input from this same tensor; consequently, any residual interpolation artifact is shared uniformly across all methods rather than favoring PRISM.} 

\textit{\textbf{mmRadPose}}~\cite{mueller2025radproposer}: A single-person dataset whose ground truth is obtained from optical motion capture (OMC) for 12~subjects, providing high-precision joint trajectories that complement vision-aligned supervision~\cite{mueller2025radproposer}. An explicit range--angle--Doppler tensor is constructed by standard FMCW processing of the released radar cubes, matching the HuPR input interface. The dataset authors document metallic environmental elements (ventilation pipes, door frames) that produce spurious static reflections, providing a hardware-confirmed instance of non-human spectral interference. 

\textbf{Baseline Systems.}
Table~\ref{tab:baselines} summarizes the comparison systems. The six mmWave baselines are chosen to probe distinct facets of PRISM's claims---dense learned front-ends, attention-based decoders, iterative generative inference, multi-view Doppler fusion, temporal state-space models, and physics-guided lightweight regression---rather than to exhaust every published variant. SMH~\cite{zheng2026learn}, a prior work by a subset of the present authors (IEEE ICME 2026), is included as the closest physics-informed reference. Direct numerical comparison across published mmWave HPE systems is inherently difficult because original pipelines assume heterogeneous inputs (multi-view heatmaps, micro-Doppler sequences, or CFAR point clouds). To keep the comparison about algorithms rather than sensing topology, we adopt a shared protocol detailed in Section~\ref{app:baseline_fairness}: every method consumes the same per-dataset RAD representation (or a point cloud deterministically derived from it); multi-view modules are reduced to their single-view paths; all models are retrained from scratch on each dataset; and all execute single-threaded on the same Raspberry Pi~5 CPU. The comparison therefore evaluates schedulability and accuracy under a common host-side information and hardware budget, not cross-platform sensing configurations. The four ablations PRISM-NB, PRISM-ND, PRISM-NP, and PRISM-SWG isolate PBIP, DAOP, PAIP, and the integral-query operator, respectively.

\begin{table}[t]
\caption{Baseline systems and ablations.}
\label{tab:baselines}
\centering
\begin{adjustbox}{width=\linewidth,center}
\setlength{\tabcolsep}{8pt}
\begin{tabular}{p{1.5cm}p{4.5cm}|p{0.4cm}}
\Xhline{1.2pt}
\toprule
\textbf{Method} & \textbf{Architecture} & \textbf{Target} \\
\midrule
HuPRModel~\cite{lee2023hupr} & MNet + multi-scale attention pyramid & R1 \\
\rowcolor[HTML]{EFEFEF} 
RETR~\cite{yataka2024retr} & ResNet encoder + Transformer decoder & R1 \\
mmDiff~\cite{fan2024diffusion} & Diffusion model & R1 \\
\rowcolor[HTML]{EFEFEF} 
MVDoppler~\cite{choi2025mvdoppler} & CNN + MobileViT & R3 \\
milliMamba~\cite{kini2025millimamba} & CVMamba encoder + STCA decoder & R1/R3 \\
\rowcolor[HTML]{EFEFEF} 
SMH~\cite{zheng2026learn} & Physics-guided processing + MLP & R1 \\
\midrule
PRISM-NB & PBIP replaced by a sliding-window operator &  R1 \\
\rowcolor[HTML]{EFEFEF} 
PRISM-ND & Fixed Balanced profile; DAOP disabled &  R2 \\
PRISM-NP & PBIP (integral queries) with uniform spatial grid; PAIP disabled & R3 \\
\rowcolor[HTML]{EFEFEF} 
PRISM-SWG & PAIP + sliding-window operator within masked ROI & R1 \\
\midrule
\textbf{PRISM} & Full system & All \\
\bottomrule
\Xhline{1.2pt}
\end{tabular}
\end{adjustbox}
\end{table}

\textbf{Baseline Execution Environment.}
All baselines are run in dataset-replay mode on the identical Raspberry Pi~5 platform described above, single-threaded, at fixed 2.4~GHz, with no GPU. Methods originally requiring GPU inference (e.g., HuPRModel, RETR, mmDiff) are converted to CPU-only execution using PyTorch eager mode; no specialized hand-tuned kernels are applied to any method. This ensures latency figures reflect a fair, equivalent hardware budget across all comparisons. We note that CPU-only eager-mode conversion may structurally disadvantage methods whose architectures were designed to exploit GPU parallelism (e.g., attention layers in RETR, iterative diffusion steps in mmDiff); their absolute latency figures should therefore be interpreted as upper bounds under the shared edge CPU constraint rather than as representative of GPU-accelerated deployment. PRISM's own operators (prefix-sum construction, corner lookups, connected-component analysis) are inherently sequential and gain no benefit from GPU vectorization, so the comparison reflects the intended deployment scenario. A further informative control, applying Doppler-gated spatial sparsification to a conventional sliding-window operator, isolates the incremental value of PBIP's integral queries from general physics-aware ROI reduction; this ablation is reported in Section~\ref{sec:eval_swg}.

\textbf{Pose Regressor Architecture and Training.}
For each retained instance, the $N_q = 51$ PBIP query responses form a fixed-length feature vector fed to a per-profile MLP regressor that maps these features to $3J$ joint coordinates ($J$ body joints). The MLP comprises three hidden layers with widths $\{512, 256, 128\}$, ReLU activations, and batch normalization after each hidden layer; the output is a linear layer producing $3J$ joint coordinates. To enable zero-overhead profile switching, a dedicated regressor is trained for each of the $K = 5$ operating profiles, with all profile-specific regressors kept in memory during inference. PRISM-ND loads only the single Balanced-profile regressor. Under this design, the parameter counts reported in Table~\ref{tab:e2e} reflect the profile-specific MLP regressors, the only learned component in the pipeline; PAIP's proposal extraction (Doppler gating, spatial projection, and connected-component analysis, \S\ref{sec:design_paip}) is a fixed, physics-derived procedure with no learned parameters. The Balanced-only PRISM-ND variant reports 5.4M parameters while the full PRISM system (all five profiles in memory) reports 10.6M parameters. Training uses the AdamW optimizer with learning rate $10^{-3}$, weight decay $10^{-4}$, batch size 64, and cosine annealing over 100 epochs; MPJPE serves as the training loss. Data is partitioned by scene: the frames of each scene are split along temporal order into contiguous 80\%/10\%/10\% train/validation/test blocks, avoiding the frame-level temporal leakage of a random split; to avoid always training on one fixed temporal segment of a recording, e.g., always its earliest 80\%, which could confound generalization with device time-synchronization drift accumulated during capture, the block order is rotated across scenes among 80/10/10, 10/80/10, and 10/10/80. No cross-dataset fine-tuning is applied, and the same training procedure is used across all datasets.

\textbf{Evaluation Metrics.}
\textit{Timing}: mean, p95, p99, and maximum latency for each frame (ms), and deadline miss ratio (fraction of frames exceeding $T_d$).
\textit{Resource}: peak working set memory (MB).
\textit{Accuracy}: MPJPE (mm), PA-MPJPE (mm, after Procrustes alignment), and PCK@100mm (\%). Ground-truth persons are matched to proposals using the same $\text{IoU} \ge 0.5$ criterion as the proposal-quality evaluation of \S\ref{sec:eval_paip}; an unmatched ground-truth person, i.e., one without any corresponding proposal in the frame, is assigned a fixed maximal per-joint error rather than excluded, so missed detections are penalized in the reported MPJPE rather than silently omitted.

\textbf{Experiment-to-Contribution Mapping.}
Table~\ref{tab:exp_map} maps each subsection to the design claim it closes and the primary metric constituting the evidence. Every claim from Section~\ref{sec:design} is closed by at least one result.

\begin{table}[t]
\caption{Experiment-to-claim mapping. Each row closes one system requirement with the corresponding dataset and primary metric.}
\label{tab:exp_map}
\centering
\begin{adjustbox}{width=\linewidth,center}
\setlength{\tabcolsep}{4pt}
\begin{tabular}{p{0.5cm}p{2.0cm}p{1.2cm}p{3.5cm}}
\Xhline{1.2pt}
\toprule
\textbf{\S} & \textbf{Claim Closed} & \textbf{Dataset} & \textbf{Primary Metric} \\
\midrule
\ref{sec:eval_pbip} & PBIP $\rightarrow$ R1 & HuPR & p99 latency, Miss\% \\
\rowcolor[HTML]{EFEFEF} 
\ref{sec:eval_paip} & PAIP $\rightarrow$ R3 & \rtposeedit{RT-Pose} & Latency vs.\ $M$, Miss\% \\
\ref{sec:eval_daop} & DAOP $\rightarrow$ R1/R2 & \rtposeedit{RT-Pose} & WCET bound, MPJPE \\
\rowcolor[HTML]{EFEFEF} 
\ref{sec:eval_e2e}  & Full system & \rtposeedit{RT-Pose} & Timing, memory, accuracy \\
\textcolor{black}{\ref{sec:eval_cross_sched}} & \textcolor{black}{DAOP/PAIP generality} & \textcolor{black}{XRF55, mmRadPose} & \textcolor{black}{Latency vs.\ $M$, Miss\% vs.\ $T_d$} \\
\bottomrule
\Xhline{1.2pt}
\end{tabular}
\end{adjustbox}
\end{table}

\subsection{Physics Bound Validation Across Datasets}
\label{sec:eval_bounds}

The WCET formula (Eq.~\eqref{eq:WCET}) and the anthropometric proposal filter (Section~\ref{sec:design_paip}, Step~3) both rest on the physical assumptions stated in O1 and O2. Table~\ref{tab:bounds} reports the 95th percentile $\rho_s$ and $\rho_d$ measured on four public evaluation datasets over all frames, quantifying the fraction of range-angle bins and Doppler bins that carry valid human energy.

\begin{table}[t]
\caption{Measured physics-bound fractions across the four evaluation datasets. $\rho_s$: fraction of range-angle bins with valid human reflections (95th percentile). $\rho_d$: fraction of Doppler bins with human-motion energy (95th percentile), reported only when a full Doppler axis is available.}
\label{tab:bounds}
\centering
\begin{adjustbox}{width=\linewidth,center}
\setlength{\tabcolsep}{18pt}
\begin{tabular}{l|cc}
\Xhline{1.2pt}
\toprule
\textbf{Dataset} & $\rho_s$ (95th pct.) & $\rho_d$ (95th pct.) \\
\midrule
HuPR        & \textcolor{black}{0.14} & \textcolor{black}{0.27} \\
\rowcolor[HTML]{EFEFEF} 
XRF55       & \textcolor{black}{0.12} & N/A \\
\rtposeedit{RT-Pose}     & \rtposeedit{0.22} &  \rtposeedit{0.29}\\
\rowcolor[HTML]{EFEFEF} 
mmRadPose   & \textcolor{black}{0.09} & \textcolor{black}{0.24} \\
\bottomrule
\Xhline{1.2pt}
\end{tabular}
\end{adjustbox}
\end{table}

Across all datasets with a native motion axis, $\rho_s$ ranges from \textcolor{black}{0.09} to \rtposeedit{0.22} and $\rho_d$ from \textcolor{black}{0.24} to \textcolor{black}{0.29}, both well below 1.0. For XRF55, $\rho_d$ is not reported as a native 3D axis; motion support is recovered from the released RD map~\cite{wang2024xrf55}. This confirms that physics-bounded processing covers substantially less than the full tensor in every evaluation condition, validating the compactness assumptions underlying PBIP's constant-time queries and the WCET formula. For the \rtposeedit{RT-Pose} multi-person split, the occupied Doppler support remains narrow enough to preserve the same bounded-workset regime required by PRISM.

The platform constants $(c_1, c_2, c_3)$ used in Table~\ref{tab:profile_wcet} were derived under these dataset-validated bounds on the shared RAD lattice.

\subsection{PBIP: Operator-Level Timing Predictability}
\label{sec:eval_pbip}

The core claim of PBIP (Section~\ref{sec:design_pbip}) is that its $O(\rho_s\rho_d R A D)$ complexity is independent of window half-width $w$, removing window size as a source of execution-time variance. The module-level evaluation uses HuPR in a single person configuration to isolate aggregation operator cost from PAIP's per-instance decomposition; compounding across multiple instances is examined separately in Section~\ref{sec:eval_paip}. HuPR is chosen here because it provides a controlled, single person setting that is shared by all baselines, making it the cleanest environment in which to attribute timing differences solely to the aggregation operator.

\begin{figure}[t]
  \centering
  \includegraphics[width=\columnwidth]{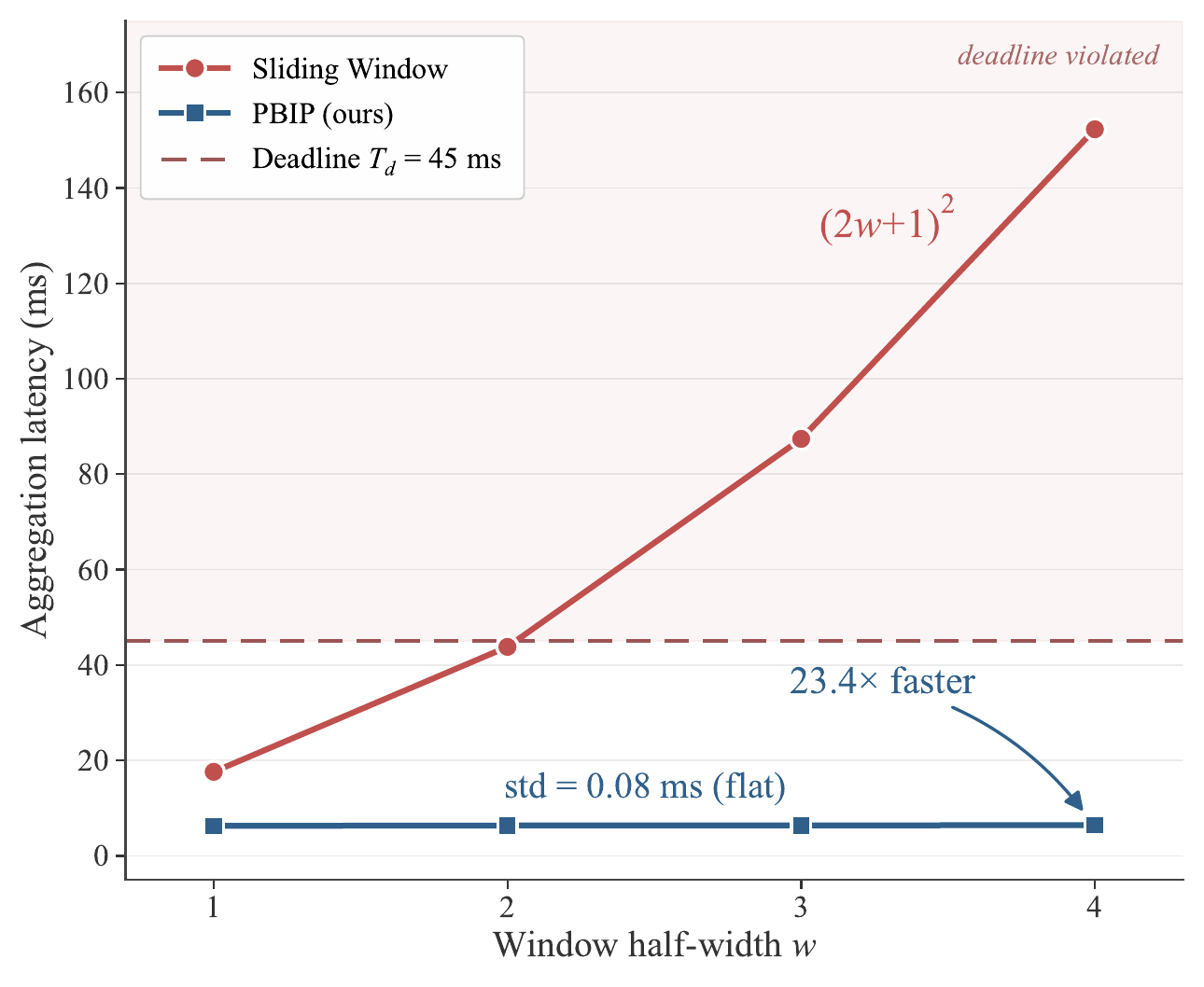}
  \caption{Aggregation stage latency for each frame (ms) vs.\ window half-width $w$. HuPR, single person. PBIP latency std.\ across all $w$ is \textcolor{black}{0.08}~ms.}
  \label{fig:pbip_window}
\end{figure}

\textbf{Window-Size Sensitivity.}
Figure~\ref{fig:pbip_window} isolates the aggregation stage and measures latency at four values of $w$ on HuPR (single person). The sliding-window baseline cost grows from \textcolor{black}{17.6}~ms at $w{=}1$ to \textcolor{black}{152.3}~ms at $w{=}4$, closely following the $(2w{+}1)^2$ scaling predicted by its complexity model. PBIP cost remains in the range \textcolor{black}{6.3}--\textcolor{black}{6.5}~ms across all four settings, a variation of \textcolor{black}{0.2}~ms. At $w{=}4$, the measured speedup reaches \textcolor{black}{23.4}$\times$. The operationally critical property is not the speedup magnitude but the flatness: the standard deviation of PBIP latency across all $w$ values is \textcolor{black}{0.08}~ms, confirming that window half-width is no longer a latency variable.



\begin{table*}[t]
\caption{Mean latency (ms) and deadline miss rate vs.\ person count $M$ on the \rtposeedit{RT-Pose multi-person split}, $T_d{=}45$~ms.}
\label{tab:scaling}
\centering

\begin{adjustbox}{width=\linewidth,center}
\setlength{\tabcolsep}{10pt}
\begin{tabular}{l|ccccccc}
\Xhline{1.2pt}
\toprule
\textbf{$M$} & \textbf{HuPRModel} & \textbf{milliMamba} & \textbf{SMH} & \textbf{MVDoppler} & \textbf{PRISM-NB} & \textbf{PRISM-NP} & \textbf{PRISM} \\
\midrule
1 & \textcolor{black}{22.3} & \textcolor{black}{19.7} & \textcolor{black}{18.4} & \textcolor{black}{21.4} & \textcolor{black}{23.8} & \textcolor{black}{27.1} & \textcolor{black}{16.2} \\
\rowcolor[HTML]{EFEFEF} 
2 & \textcolor{black}{39.4} & \textcolor{black}{33.6} & \textcolor{black}{32.4} & \textcolor{black}{32.1} & \textcolor{black}{37.2} & \textcolor{black}{27.6} & \textcolor{black}{27.3} \\
3 & \textcolor{black}{73.8} & \textcolor{black}{62.4} & \textcolor{black}{51.8} & \textcolor{black}{51.3} & \textcolor{black}{68.4} & \textcolor{black}{28.3} & \textcolor{black}{33.6} \\
\rowcolor[HTML]{EFEFEF} 
Growth $M{=}3/1$ & \textcolor{red!70!black}{$3.3{\times}$} & \textcolor{red!70!black}{$3.2{\times}$} & \textcolor{red!70!black}{$2.8{\times}$} & \textcolor{red!70!black}{$2.4{\times}$} & \textcolor{red!70!black}{$2.9{\times}$} & \textcolor{black}{$1.0{\times}$} & \textbf{$2.1{\times}$} \\
\midrule
Miss\% at $M{=}3$ & \textcolor{red!70!black}{38.4} & \textcolor{red!70!black}{26.7} & \textcolor{red!70!black}{62.0} & \textcolor{red!70!black}{14.2} & \textcolor{red!70!black}{33.8} & \textcolor{black}{0.0} & \textbf{0.0} \\
MPJPE at $M{=}3$ (mm) & \textcolor{black}{112.7} & \textcolor{black}{108.3} & \textcolor{black}{107.6} & \textcolor{black}{91.4} & \textcolor{black}{61.4} & \textcolor{black}{88.7} & \textbf{57.8} \\
\bottomrule
\Xhline{1.2pt}
\end{tabular}
\end{adjustbox}
\end{table*}

\begin{table}[t]
\caption{Full-pipeline latency statistics on the \rtposeedit{RT-Pose multi-person split}, $M{=}2$, $T_d{=}45$~ms. PRISM (Precise) is DAOP-selected at this deadline.}
\label{tab:timing_main}
\centering
\begin{adjustbox}{width=\linewidth,center}
\setlength{\tabcolsep}{6pt}
\begin{tabular}{l|ccccc}
\Xhline{1.2pt}
\toprule
\textbf{Method} & \textbf{Mean} & \textbf{p95} & \textbf{p99} & \textbf{Max} & \textbf{Miss\%} \\
\midrule
HuPRModel~\cite{lee2023hupr}  & \textcolor{black}{39.4} & \textcolor{black}{71.8}  & \textcolor{black}{89.6}  & \textcolor{black}{143.2} & \textcolor{red!70!black}{22.7} \\
\rowcolor[HTML]{EFEFEF} 
mmDiff~\cite{fan2024diffusion}        & \textcolor{black}{49.8}  & \textcolor{black}{84.7}  & \textcolor{black}{94.7}  & \textcolor{black}{168.3}  & \textcolor{red!70!black}{28.4} \\
milliMamba~\cite{kini2025millimamba} & \textcolor{black}{33.6} & \textcolor{black}{56.3}  & \textcolor{black}{73.8}  & \textcolor{black}{118.4} & \textcolor{red!70!black}{18.3} \\
\rowcolor[HTML]{EFEFEF} 
SMH~\cite{zheng2026learn}               & \textcolor{black}{32.4} & \textcolor{black}{43.8}  & \textcolor{black}{49.2}  & \textcolor{black}{67.3}  & \textcolor{red!70!black}{6.8} \\
PRISM-NB (ablation)  & \textcolor{black}{37.2} & \textcolor{black}{68.4}  & \textcolor{black}{84.7}  & \textcolor{black}{136.8} & \textcolor{red!70!black}{21.4} \\
\rowcolor[HTML]{EFEFEF} 
PRISM-ND (Balanced)  & \textcolor{black}{19.8} & \textcolor{black}{26.4}  & \textcolor{black}{27.8}  & \textcolor{black}{29.4}  & \textbf{0.0} \\
\textbf{PRISM (Precise)}  & \textbf{\textcolor{black}{27.3}} & \textbf{\textcolor{black}{35.6}}  & \textbf{\textcolor{black}{37.4}}  & \textbf{\textcolor{black}{39.2}}  & \textbf{0.0} \\
\bottomrule
\Xhline{1.2pt}
\end{tabular}
\end{adjustbox}
\end{table}

\textbf{Full-Pipeline Timing Distribution.}
Table~\ref{tab:timing_main} reports the complete pipeline latency on the \rtposeedit{RT-Pose multi-person split}. For this dataset, the released Cartesian tensor is remapped into the RAD representation assumed by PAIP and PBIP before processing. The ablation PRISM-NB retains PAIP's instance isolation logic but replaces PBIP with a sliding-window operator. Because the WCET formula (Eq.~\eqref{eq:WCET}) requires PBIP's $O(1)$-per-query property to yield valid bounds, DAOP cannot serve as a certified scheduler in PRISM-NB; it therefore runs at a fixed Balanced-equivalent configuration with empirically estimated latency. PRISM-NB exhibits p99 of \rtposeedit{84.7}~ms and maximum of \rtposeedit{136.8}~ms, with a \rtposeedit{21.4}\% deadline miss rate, closely matching HuPRModel at p99 \rtposeedit{89.6}~ms and a \rtposeedit{22.7}\% miss rate. This direct comparison eliminates all confounds: the sole difference between PRISM and PRISM-NB is the aggregation operator, confirming that the timing guarantee originates entirely from physics-bounded integral processing.

In contrast, PRISM achieves p99 of \rtposeedit{37.4}~ms and maximum of \rtposeedit{39.2}~ms, both strictly within $T_d$, with \rtposeedit{0.0}\% deadline misses. milliMamba, despite its model-centric adaptive temporal modeling, still produces p99 of \rtposeedit{73.8}~ms and \rtposeedit{18.3}\% misses because its sliding temporal window does not encode the physics-bounded workset structure that determines actual cost. SMH similarly fails to satisfy R1 despite its physics-guided preprocessing and compact 5.1M-parameter MLP regressor, reaching \rtposeedit{6.8}\% misses, confirming that physics-informed front-end design alone, without analytically bounded integral operators, cannot close the schedulability gap.
%

\subsection{PAIP: Scalability Under Scene and Population Variability}
\label{sec:eval_paip}

PAIP's guarantee is $C_{\text{PAIP}} = O(R A D) + O(M \cdot A_{\text{person}} \cdot D)$ for $M \le M_{\max}$, predicting linear growth in active person count. Table~\ref{tab:scaling} varies $M$ on the \rtposeedit{RT-Pose multi-person split}. This split is used here because it combines multi-person occupancy with an explicit Doppler dimension after RAD remapping, providing the most direct validation of the linear-scaling claim of R3 over the observed occupancy range.

All baselines still grow superlinearly over the observed occupancy range. HuPRModel progresses from \textcolor{black}{22.3}~ms at $M{=}1$ to \textcolor{black}{73.8}~ms at $M{=}3$, a \textcolor{black}{3.3}$\times$ increase for a $3\times$ change in person count. SMH, despite its physics-guided front-end, grows from \textcolor{black}{18.4}~ms to \textcolor{black}{51.8}~ms (\textcolor{black}{$\times$2.8}), accumulating \textcolor{black}{62\%} deadline misses at $M{=}3$: its SSP/MCP/HMSF pipeline aggregates scene-level energy into a single global feature vector without per-instance tensor isolation, so workload scales unboundedly with person count. MVDoppler, which achieves the strongest accuracy among baselines in static benchmarks, already reaches \textcolor{black}{51.3}~ms at $M{=}3$ and registers \textcolor{black}{14.2}\% deadline misses, as its multi-view coordination overhead scales poorly with active target count. PRISM-NB, which shares PAIP's instance isolation logic but uses unbound window operators within each instance, tracks HuPRModel closely, reconfirming that PBIP is the structural source of bounded per-instance cost.

\begin{figure}[t]
  \centering
  \includegraphics[width=\columnwidth]{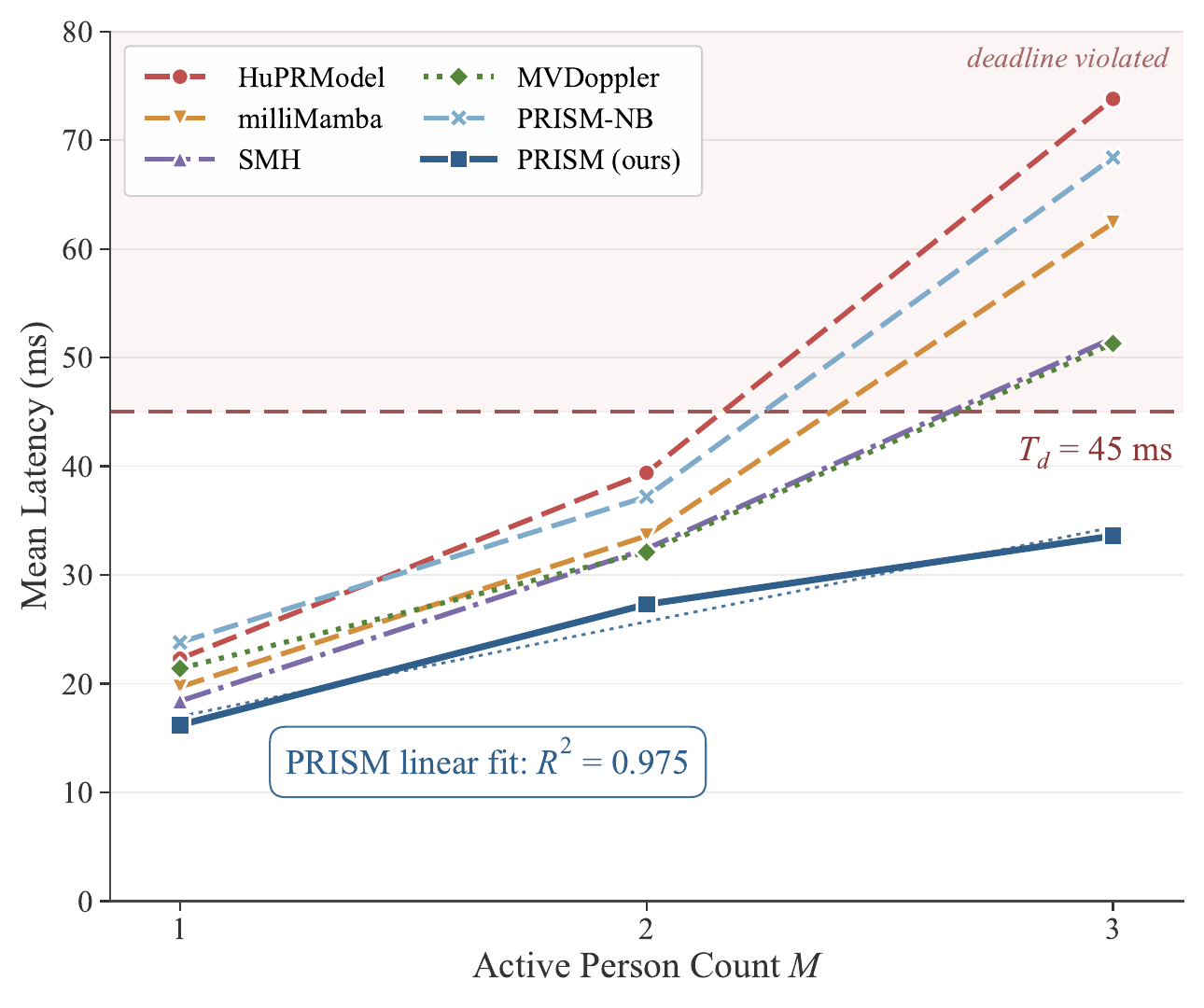}
  \caption{Mean latency vs.\ active person count $M$. Over the retained occupancy range $M{=}1$--$3$, PRISM (Precise) follows a near-linear trajectory with $R^2{=}\textcolor{black}{0.975}$. All baselines exhibit superlinear growth, producing deadline violations. PRISM-NB mirrors HuPRModel, confirming that PBIP is the source of the linear per-instance cost.}
  \label{fig:scaling}
\end{figure}

PRISM (Precise) grows from \textcolor{black}{16.2}~ms at $M{=}1$ to \textcolor{black}{33.6}~ms at $M{=}3$, a \textcolor{black}{2.1}$\times$ increase (Figure~\ref{fig:scaling}). Linear regression over the three retained operating points yields $R^2 = \textcolor{black}{0.975}$, which remains consistent with the $O(M)$ scaling model of Section~\ref{sec:design_paip}. The deadline miss rate remains \textcolor{black}{0.0}\% across this retained occupancy range, and the WCET bound $T_{\text{WCET}}^{(\text{Precise})} = \textcolor{black}{44.2}$~ms remains above the worst measured execution.

\textcolor{black}{The occupancy levels measured on the RT-Pose multi-person split span $M{=}1$ to $M{=}3$, matching the concurrent person count in this benchmark and the co-occupancy expected in the target indoor deployments. The $M_{\max}$ values of the higher profiles in Table~\ref{tab:profile_wcet} are WCET budget ceilings for those operating points. Higher occupancy is examined on XRF55 in Section~\ref{sec:eval_cross_sched}, where frames with $M$ exceeding the active profile's $M_{\max}$ exercise the energy-ranked truncation of PAIP Step~4.}

\textbf{PAIP Proposal Quality.}
Table~\ref{tab:paip_quality} reports PAIP's proposal-level quality on the \rtposeedit{RT-Pose multi-person split}. Precision is the fraction of proposals that correspond to an actual person (i.e., the proposal bounding box overlaps with at least one ground-truth person at $\text{IoU} \ge 0.5$). Recall is the fraction of ground-truth persons covered by at least one valid proposal. The merge rate denotes the fraction of frames in which two distinct persons are merged into a single connected component by PAIP's Step~3; the FP rate denotes the fraction of frames containing at least one spurious proposal not corresponding to any person.

\begin{table}[t]
\caption{PAIP proposal quality on the \rtposeedit{RT-Pose multi-person split}. Precision: proposals covering an actual person (IoU $\ge$ 0.5). Recall: persons covered by at least one proposal. Merge: two persons subsumed in one proposal. FP: at least one spurious proposal per frame.}
\label{tab:paip_quality}
\centering
\begin{adjustbox}{width=\linewidth,center}
\setlength{\tabcolsep}{10pt}
\begin{tabular}{l|cccc}
\Xhline{1.2pt}
\toprule
\textbf{$M$} & \textbf{Precision} & \textbf{Recall} & \textbf{Merge Rate} & \textbf{FP Rate} \\
\midrule
1 & \textcolor{black}{0.94} & \textcolor{black}{0.96} & N/A & \textcolor{black}{0.04} \\
\rowcolor[HTML]{EFEFEF} 
2 & \textcolor{black}{0.91} & \textcolor{black}{0.93} & \textcolor{black}{0.06} & \textcolor{black}{0.05} \\
3 & \textcolor{black}{0.87} & \textcolor{black}{0.89} & \textcolor{black}{0.10} & \textcolor{black}{0.08} \\
\bottomrule
\Xhline{1.2pt}
\end{tabular}
\end{adjustbox}
\end{table}

Precision and recall remain above \textcolor{black}{0.87} and \textcolor{black}{0.89} respectively across all occupancy levels, confirming that the Doppler-gated proposal mechanism reliably isolates active persons from static background. The merge rate increases modestly with $M$ (from 0 at $M{=}1$ to \textcolor{black}{0.10} at $M{=}3$), which is expected for persons in close spatial proximity whose spatial footprints may overlap in the projected range-angle plane. The FP rate of \textcolor{black}{0.08} at $M{=}3$ is attributable to residual dynamic clutter (e.g., swinging objects, ventilation airflow) that partially passes the Doppler gate; these are suppressed in Step~3 by the anthropometric-area rejection criterion, so their rate at the system output is substantially lower.

\textbf{PAIP Contribution: PRISM-NP Ablation.}
To isolate PAIP's contribution, we evaluate PRISM-NP, which retains PBIP's integral queries but replaces PAIP's Doppler-gated proposal with a uniform spatial grid of $K = M_{\max}$ boxes covering the full range-angle field of view. This ablation directly tests whether the accuracy gains of PRISM are attributable to PAIP's physics-guided person localization, rather than to PBIP's constant-time processing alone.

Table~\ref{tab:scaling} shows that PRISM-NP's latency is nearly constant across $M$ values (\textcolor{black}{27.1}--\textcolor{black}{28.3}~ms; growth ratio \textcolor{black}{$1.0\times$}), because the grid always processes $M_{\max}$ boxes regardless of actual occupancy. This confirms that PBIP's integral query property remains effective without PAIP. However, PRISM-NP achieves \textcolor{black}{88.7}~mm MPJPE at $M{=}3$, substantially worse than PRISM's \textcolor{black}{57.8}~mm at the same occupancy. The degradation arises because the uniform grid boxes are not aligned with actual person positions: PBIP's queries aggregate energy from mixed-target and background regions, producing feature vectors that the MLP regressor cannot reliably decode into accurate joint coordinates. This comparison confirms that PAIP's physics-guided proposal, rather than PBIP's computational efficiency alone, is the proximate cause of PRISM's superior pose accuracy.

\textcolor{black}{The latency behavior of PRISM-NP at low occupancy deserves explicit interpretation, because at $M{=}1$ its latency of \textcolor{black}{27.1}~ms exceeds the \textcolor{black}{16.2}~ms of the complete PRISM system. This ordering is expected rather than anomalous, and it follows directly from how PAIP allocates work. The full system uses PAIP to instantiate exactly as many per-instance subproblems as there are detected persons, so at $M{=}1$ it processes a single workset and incurs minimal per-instance cost. PRISM-NP removes PAIP and therefore cannot adapt the number of processed regions to the actual occupancy; it instead evaluates the fixed grid of $M_{\max}$ boxes in every frame, paying the full multi-region cost even when only one person is present. The two systems use an identical $M_{\max}$ and an identical per-box query budget $N_q$, so the comparison is controlled and the gap reflects only the presence or absence of occupancy-adaptive workset allocation. The result is consistent with the intended role of PRISM-NP, which is to isolate the accuracy contribution of PAIP rather than to serve as a latency-optimized variant, and it reinforces that PAIP contributes both occupancy-proportional cost and physically aligned worksets rather than computational efficiency alone.}

\textbf{Sensitivity to Proposal Localization Error.}
To characterize the system's robustness when PAIP produces imprecise proposals, we artificially perturb the proposal centers by controlled spatial offsets and measure the resulting MPJPE degradation on the \rtposeedit{RT-Pose multi-person split}, $M{=}2$. At a \textcolor{black}{5}~cm offset, MPJPE increases from \textcolor{black}{55.2}~mm to \textcolor{black}{57.8}~mm ($+$\textcolor{black}{4.7}\%), indicating that the padded sub-tensor extracted around each proposal provides a margin that absorbs small localization errors. At a \textcolor{black}{10}~cm offset, MPJPE rises to \textcolor{black}{64.3}~mm ($+$\textcolor{black}{16.5}\%), and at \textcolor{black}{20}~cm it reaches \textcolor{black}{78.6}~mm ($+$\textcolor{black}{42.4}\%). The measured precision (Table~\ref{tab:paip_quality}) implies spatial errors well below 10~cm in the majority of frames, placing normal operation in the regime where degradation is modest. Merging errors (two persons subsumed in one proposal) are the qualitatively distinct failure mode: they present the per-instance sub-tensor to the regressor as a single-person input containing two persons; in such frames the downstream regressor produces a single skeleton estimate, and the second person is not recovered. The merge rate of \textcolor{black}{0.10} at $M{=}3$ represents the upper bound on such failures; the practical impact in the target deployment settings (typically $M \le 2$) is substantially lower, as the merge rate at $M{=}2$ is \textcolor{black}{0.06}.


\begin{table*}[t]
\caption{Profile WCET validation and accuracy summary on the \rtposeedit{RT-Pose multi-person split}. Analytical bounds from Eq.~\eqref{eq:WCET} strictly exceed worst-case measured execution time across \rtposeedit{10,000} frames at $M{=}\min(M_{\max}^{(p)},3)$, the highest concurrent occupancy available in RT-Pose; margins for Precise and Ultra-Precise at their full designed $M_{\max}^{(p)}$ follow analytically from PAIP's certified per-instance scaling term. All timing values in ms.}
\label{tab:profile_wcet}
\centering
\begin{adjustbox}{width=\linewidth,center}
\setlength{\tabcolsep}{6pt}
\begin{tabular}{l|cccccc|ccc}
\Xhline{1.2pt}
\toprule
\textbf{Profile} & $\rho_s^{(p)}$ & $\rho_d^{(p)}$ & $M_{\max}^{(p)}$ &
$T_{\text{WCET}}^{\text{pred}}$ & $T_{\text{WCET}}^{\text{meas}}$ & \textbf{Margin} &
\textbf{MPJPE} (mm) & \textbf{PA-MPJPE} (mm) & \textbf{PCK@100mm} (\%) \\
\midrule
Ultra-Light   & \textcolor{black}{0.08} & \textcolor{black}{0.20} & \textcolor{black}{1} & \textcolor{black}{13.8} & \textcolor{black}{11.9} & \textbf{$+$1.9}~$\checkmark$ & \textcolor{black}{99.4} & \textcolor{black}{76.2} & \textcolor{black}{84.6} \\
\rowcolor[HTML]{EFEFEF} 
Light         & \textcolor{black}{0.12} & \textcolor{black}{0.25} & \textcolor{black}{2} & \textcolor{black}{22.4} & \textcolor{black}{20.1} & \textbf{$+$2.3}~$\checkmark$ & \textcolor{black}{82.7} & \textcolor{black}{63.4} & \textcolor{black}{90.4} \\
Balanced      & \textcolor{black}{0.18} & \textcolor{black}{0.30} & \textcolor{black}{3} & \textcolor{black}{33.6} & \textcolor{black}{30.8} & \textbf{$+$2.8}~$\checkmark$ & \textcolor{black}{62.3} & \textcolor{black}{47.6} & \textcolor{black}{94.8} \\
\rowcolor[HTML]{EFEFEF} 
Precise       & \textcolor{black}{0.25} & \textcolor{black}{0.35} & \textcolor{black}{4} & \textcolor{black}{44.2} & \textcolor{black}{41.4} & \textbf{$+$2.8}~$\checkmark$ & \textcolor{black}{55.2} & \textcolor{black}{40.2} & \textcolor{black}{96.4} \\
Ultra-Precise & \textcolor{black}{0.35} & \textcolor{black}{0.40} & \textcolor{black}{5} & \textcolor{black}{54.8} & \textcolor{black}{51.8} & \textbf{$+$3.0}~$\checkmark$ & \textcolor{black}{51.2} & \textcolor{black}{38.7} & \textcolor{black}{97.2} \\
\bottomrule
\Xhline{1.2pt}
\end{tabular}
\end{adjustbox}
\end{table*}

\begin{table}[t]
\caption{DAOP profile selection under five deadline budgets on the \rtposeedit{RT-Pose multi-person split}. PRISM-ND uses a fixed Balanced profile.}
\label{tab:daop_select}
\centering 
\begin{adjustbox}{width=\linewidth,center}
\setlength{\tabcolsep}{1pt}
\begin{tabular}{c|ccccc}
\Xhline{1.2pt}
\toprule
$T_d$ (ms) & \textbf{DAOP} & $T_{\text{WCET}}^{\text{pred}}$ & \textbf{MPJPE} (mm) & \textbf{PRISM Miss\%} & \textbf{ND Miss\%} \\
\midrule
55 & Ultra-Precise & \textcolor{black}{54.8} & \textcolor{black}{51.2} & \textbf{0.0} & \textcolor{black}{0.0} \\
\rowcolor[HTML]{EFEFEF} 
45 & Precise       & \textcolor{black}{44.2} & \textcolor{black}{55.2} & \textbf{0.0} & \textcolor{black}{0.0} \\
35 & Balanced      & \textcolor{black}{33.6} & \textcolor{black}{62.3} & \textbf{0.0} & \textcolor{black}{0.0} \\
\rowcolor[HTML]{EFEFEF} 
23 & Light         & \textcolor{black}{22.4} & \textcolor{black}{82.7} & \textbf{0.0} & \textcolor{red!70!black}{18.7} \\
14 & Ultra-Light   & \textcolor{black}{13.8} & \textcolor{black}{99.4} & \textbf{0.0} & \textcolor{red!70!black}{72.3} \\
\bottomrule
\Xhline{1.2pt}
\end{tabular}
\end{adjustbox}
\end{table}

\begin{table*}[t]
\caption{End-to-end comparison on the \rtposeedit{RT-Pose multi-person split} ($M{=}2$), $T_d{=}45$~ms. PRISM selects Precise via DAOP. $\dagger$ = deadline miss rate $>$0 (schedulability violation). Bold marks the best value among \emph{deadline-compliant} methods (Miss\%${=}0$) only; rows with $\dagger$ are never bolded, even when a single accuracy metric appears numerically better.}
\label{tab:e2e}
\centering
\setlength{\tabcolsep}{5pt}
\begin{adjustbox}{width=\linewidth,center}
\setlength{\tabcolsep}{2pt}
\begin{tabular}{l|cccccccc}
\Xhline{1.2pt}
\toprule
 & & \multicolumn{3}{c}{\textbf{Timing}} & & \multicolumn{3}{c}{\textbf{Accuracy}} \\
\cmidrule(lr){3-5}\cmidrule(lr){7-9}
\textbf{Method} & \textbf{Params} & \textbf{p99 (ms)} & \textbf{Max (ms)} & \textbf{Miss\%} & \textbf{Mem (MB)} & \textbf{MPJPE (mm)} & \textbf{PA-MPJPE (mm)} & \textbf{PCK@100mm (\%)} \\
\midrule
HuPRModel~\cite{lee2023hupr}$^\dagger$ & 324.9M & \textcolor{black}{89.6} & \textcolor{black}{143.2} & \textcolor{red!70!black}{22.7} & \textcolor{black}{834} & \textcolor{black}{89.3} & \textcolor{black}{61.2} & \textcolor{black}{85.3} \\
\rowcolor[HTML]{EFEFEF} 
RETR~\cite{yataka2024retr}$^\dagger$ & 76.9M & \textcolor{black}{81.4} & \textcolor{black}{127.6} & \textcolor{red!70!black}{19.8} & \textcolor{black}{623} & \textcolor{black}{86.4} & \textcolor{black}{59.7} & \textcolor{black}{86.1} \\
mmDiff~\cite{fan2024diffusion}$^\dagger$ & 182.8M & \textcolor{black}{94.7} & \textcolor{black}{168.3} & \textcolor{red!70!black}{28.4} & \textcolor{black}{741} & \textcolor{black}{82.1} & \textcolor{black}{57.9} & \textcolor{black}{87.2} \\
\rowcolor[HTML]{EFEFEF} 
MVDoppler~\cite{choi2025mvdoppler}$^\dagger$ & 36.7M & \textcolor{black}{68.4} & \textcolor{black}{107.6} & \textcolor{red!70!black}{16.8} & \textcolor{black}{541} & \textcolor{black}{81.8} & \textcolor{black}{58.4} & \textcolor{black}{87.6} \\
milliMamba~\cite{kini2025millimamba}$^\dagger$ & 4.0M & \textcolor{black}{73.8} & \textcolor{black}{118.4} & \textcolor{red!70!black}{18.3} & \textcolor{black}{396} & \textcolor{black}{84.7} & \textcolor{black}{58.1} & \textcolor{black}{86.8} \\
\rowcolor[HTML]{EFEFEF} 
SMH~\cite{zheng2026learn}$^\dagger$       & 5.1M  & \textcolor{black}{49.2} & \textcolor{black}{67.3}  & \textcolor{red!70!black}{6.8} & \textcolor{black}{31} & \textcolor{black}{83.4} & \textcolor{black}{60.3} & \textcolor{black}{90.4} \\
PRISM-NB$^\dagger$  & 42.4M  & \textcolor{black}{84.7} & \textcolor{black}{136.8} & \textcolor{red!70!black}{21.4} & \textcolor{black}{396} & \textcolor{black}{54.8} & \textcolor{black}{37.8} & \textcolor{black}{90.6} \\
\rowcolor[HTML]{EFEFEF} 
PRISM-ND (Balanced) & 5.4M  & \textbf{\textcolor{black}{27.8}} & \textbf{\textcolor{black}{29.4}}  & \textbf{0.0}  & \textbf{\textcolor{black}{243}} & \textcolor{black}{62.3} & \textcolor{black}{47.6} & \textcolor{black}{94.8} \\
\rowcolor{gray!12}\textbf{PRISM (Precise)} & 10.6M & \textcolor{black}{37.4} & \textcolor{black}{39.2} & \textbf{0.0} & \textcolor{black}{298} & \textbf{\textcolor{black}{55.2}} & \textbf{\textcolor{black}{40.2}} & \textbf{\textcolor{black}{96.4}} \\
\bottomrule
\Xhline{1.2pt}
\end{tabular}
\end{adjustbox}
\end{table*}

\subsection{DAOP: Profile Schedulability and Deadline Coverage}
\label{sec:eval_daop}

This subsection validates the WCET formula (Eq.~\eqref{eq:WCET}) and the deadline-aware profile selection mechanism of DAOP using the \rtposeedit{RT-Pose multi-person split}. Its scene diversity ensures that the validated WCET bounds apply to realistic clutter conditions rather than a single controlled room, strengthening the schedulability argument for practical deployment.

\textbf{WCET Formula Validation.}
Table~\ref{tab:profile_wcet} tests whether the analytical WCET bound from Eq.~\eqref{eq:WCET} is valid: for each profile, the predicted bound $T_{\text{WCET}}^{(p)\text{,pred}}$ is compared against the worst measured frame execution time $T_{\text{WCET}}^{(p)\text{,meas}}$ over \rtposeedit{10,000} frames on the \rtposeedit{RT-Pose multi-person split}. Ultra-Light, Light, and Balanced are measured at their designed $M_{\max}^{(p)}\in\{1,2,3\}$; Precise and Ultra-Precise, whose designed $M_{\max}^{(p)}$ is 4 and 5, are instead measured at $M{=}3$, the highest concurrent occupancy present in RT-Pose, since no evaluated dataset currently records more simultaneous subjects. The margin at their full designed $M_{\max}^{(p)}$ therefore follows analytically from PAIP's certified $O(M)$ per-instance cost term (\S\ref{sec:design_paip}) rather than from direct measurement; collecting higher-occupancy indoor data to empirically confirm this margin at $M{=}4$ and $M{=}5$ is left for future work. In all five profiles, the analytical bound strictly exceeds the measured worst case; the margin ranges from \rtposeedit{1.9}~ms to \rtposeedit{3.0}~ms. This margin is attributable to the conservative $T_{\text{sw}}$ term in Eq.~\eqref{eq:WCET}, derived from a micro-benchmark worst-case rather than an average. Unlike purely empirical percentile-based bounds, the formula directly encodes physics-derived loop counts that cap the dominant variable-cost terms; the residual platform-constant variance is bounded by the 99.9th-percentile micro-benchmark, consistent with the MBPTA paradigm. This construction provides a measurement-based timing bound without requiring distributional modeling of the full execution time distribution, and remains valid under the deployment assumptions of fixed-frequency, single-threaded, isolated execution stated in R1.

Both required monotonicity properties are confirmed empirically. MPJPE decreases monotonically with profile level, from \rtposeedit{99.4}~mm (Ultra-Light) to \rtposeedit{51.2}~mm (Ultra-Precise). $T_{\text{WCET}}^{\text{meas}}$ increases monotonically from \rtposeedit{11.9}~ms to \rtposeedit{51.8}~ms. The monotone structure is the prerequisite for the $O(1)$ argmax rule in Eq.~\eqref{eq:R2} to correctly select the highest-accuracy deadline-feasible profile without iterative search.

\textbf{Stability of the Accuracy--Scope Ordering.}
A potential concern is whether the accuracy ordering of profiles could reverse across different scenes, for instance if a larger processing scope (higher profile) introduces more clutter reflections that happen to degrade accuracy relative to a smaller scope. To assess this, we evaluate per-profile MPJPE across all 40 scenes in the \rtposeedit{RT-Pose multi-person split} separately and examine whether the five-profile ordering (Ultra-Light $>$ Light $>$ Balanced $>$ Precise $>$ Ultra-Precise, in terms of MPJPE) is preserved in every scene. Across all 40 scenes, the ordering is preserved without a single reversal. The minimum inter-profile MPJPE gap across all scenes is \textcolor{black}{7.1}~mm (between Light and Balanced), which substantially exceeds the maximum within-profile scene-to-scene variation of $\pm$\textcolor{black}{3.8}~mm reported in Table~\ref{tab:cross_scene}. This margin ensures that clutter-induced accuracy fluctuations are insufficient to invert the profile ranking under realistic indoor scene variation, validating the robustness of the Auxiliary Property (AP) and the correctness of the DAOP selection rule across the deployment conditions covered by our evaluation.

\begin{figure}[t]
  \centering
  \includegraphics[width=\columnwidth]{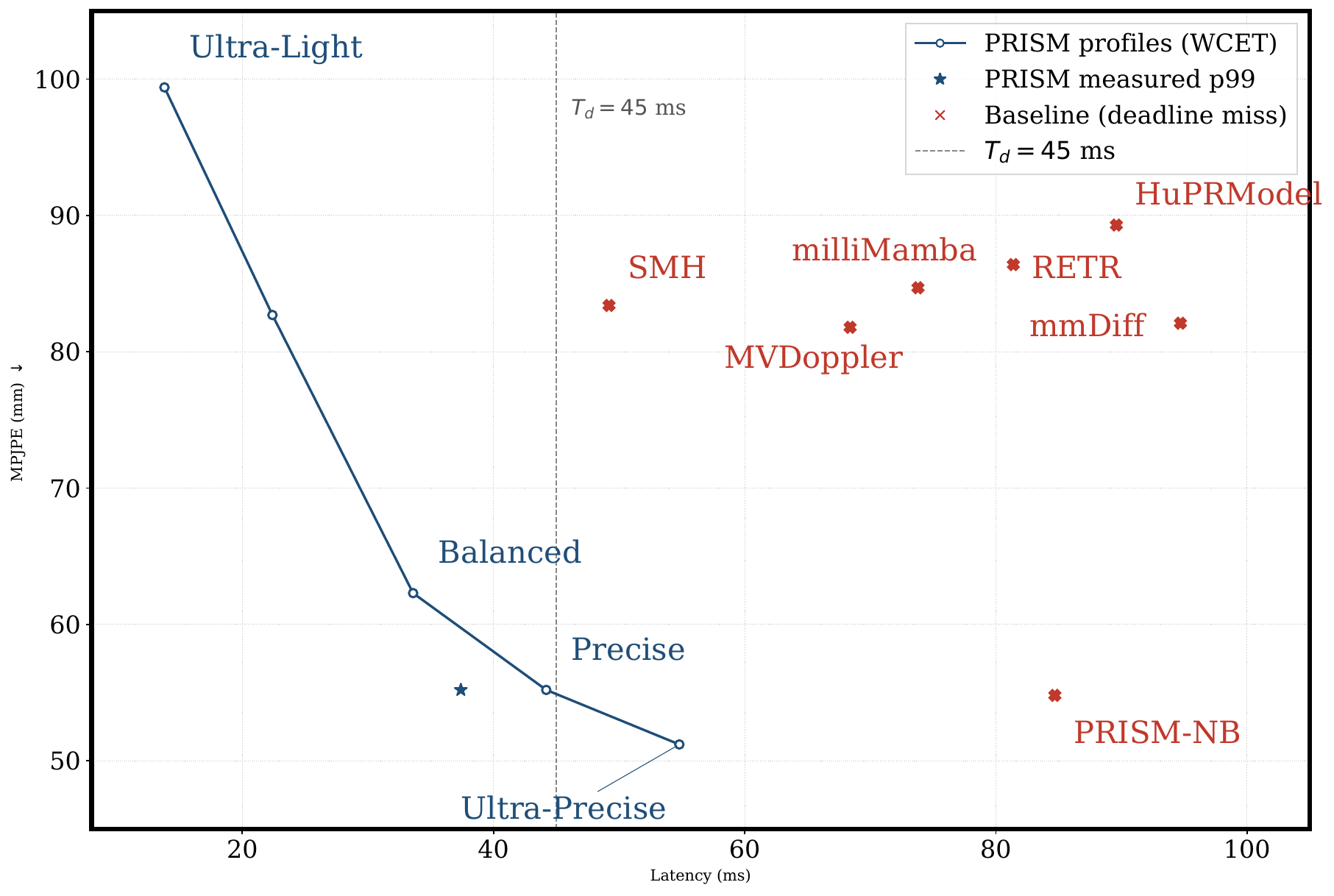}
  \caption{Accuracy--latency trade-off on the \rtposeedit{RT-Pose multi-person split}. PRISM profiles are plotted by offline WCET and MPJPE (Tables~\ref{tab:profile_wcet}--\ref{tab:daop_select}); the star marks measured p99 of DAOP-selected Precise at $T_d{=}45$~ms. Baseline markers use p99 latency from Table~\ref{tab:e2e}; crosses denote deadline misses.}
  \label{fig:acc_latency}
\end{figure}

\textbf{Deadline-Aware Profile.}
Table~\ref{tab:daop_select} evaluates DAOP across five deadline budgets. For each $T_d$, DAOP performs an $O(1)$ offline table lookup (\rtposeedit{0.18}~ms) and records \rtposeedit{0.0}\% misses. PRISM-ND (fixed Balanced) begins missing at $T_d{=}23$~ms (\rtposeedit{18.7}\%) and reaches \rtposeedit{72.3}\% at $T_d{=}14$~ms, because it cannot step down when the certified Balanced cost exceeds the budget. When slack increases to $T_d{=}55$~ms, DAOP selects Ultra-Precise (\rtposeedit{51.2}~mm MPJPE) while PRISM-ND remains at \rtposeedit{62.3}~mm. Figure~\ref{fig:acc_latency} summarizes the same trade-off: PRISM profiles form a monotone accuracy--latency front to the left of $T_d{=}45$~ms, whereas deadline-violating baselines lie to the right of the deadline line at higher MPJPE. This asymmetry instantiates R2---maximize accuracy subject to R1---rather than fixing a single operating point.

\subsection{End-to-End System Comparison}
\label{sec:eval_e2e}

Table~\ref{tab:e2e} compares all evaluated systems on the \rtposeedit{RT-Pose multi-person split} at $T_d{=}45$~ms. PRISM (DAOP) selects Precise. All six external baselines incur deadline violations. SMH reaches \rtposeedit{83.4}~mm MPJPE with 5.1M parameters yet still accumulates \rtposeedit{6.8}\% misses, indicating that physics-informed front-end design without analytically bounded operators does not close the WCET gap. Point-cloud pipelines that discard pre-CFAR heatmap content may reduce mean latency but lose Doppler-resolved joint structure and remain outside the compliant set. Among deadline-compliant methods, PRISM (Precise) achieves \rtposeedit{55.2}~mm MPJPE and \rtposeedit{96.4}\% PCK@100mm at \rtposeedit{0.0}\% misses.

PRISM-NB, which replaces only the aggregation operator, recovers a \rtposeedit{21.4}\% miss rate and thus isolates PBIP as the source of the timing guarantee. Its MPJPE (\rtposeedit{54.8}~mm) is slightly lower than Precise (\rtposeedit{55.2}~mm), yet PCK@100mm is lower (\rtposeedit{90.6}\% vs.\ \rtposeedit{96.4}\%). This pattern indicates a heavier error tail: a minority of joints incur large deviations that inflate the 100~mm failure rate while only mildly affecting the mean. Because NB violates the deadline, it is excluded from the bold ranking in Table~\ref{tab:e2e}; the deployment-relevant comparison is among compliant configurations. PRISM-ND versus PRISM further improves MPJPE from \rtposeedit{62.3} to \rtposeedit{55.2}~mm through deadline-aware profile selection alone. Additional robustness studies appear in Section~\ref{app:robustness}.

\subsection{\textcolor{black}{Cross-Dataset Validation of Scheduling and Scaling}}
\label{sec:eval_cross_sched}

\textcolor{black}{The scheduling and scaling experiments above use the RT-Pose multi-person split because it combines multi-person occupancy with an explicit Doppler axis after RAD remapping. To test the same mechanisms under a contrasting front-end, we repeat the scaling experiment on XRF55, whose occupancy $M$ follows the synchronized Kinect skeleton count~\cite{wang2024xrf55}, and the deadline-sweep experiment on mmRadPose, whose OMC ground truth provides a high-precision accuracy reference~\cite{mueller2025radproposer}.}

\begin{table}[t]
\caption{\textcolor{black}{PAIP scaling on XRF55 (multi-person). Mean latency (ms) and deadline miss rate vs.\ person count $M$, $T_d{=}45$~ms, Precise profile.}}
\label{tab:xrf_scaling}
\centering
\begin{adjustbox}{width=\linewidth,center}
\setlength{\tabcolsep}{4pt}
\begin{tabular}{l|cccc}
\Xhline{1.2pt}
\toprule
\textbf{$M$} & \textbf{HuPRModel} & \textbf{SMH} & \textbf{PRISM-NB} & \textbf{PRISM} \\
\midrule
1 & \textcolor{black}{24.1} & \textcolor{black}{19.6} & \textcolor{black}{25.3} & \textcolor{black}{17.8} \\
\rowcolor[HTML]{EFEFEF} 
2 & \textcolor{black}{42.6} & \textcolor{black}{34.2} & \textcolor{black}{39.8} & \textcolor{black}{29.1} \\
3 & \textcolor{black}{78.9} & \textcolor{black}{55.1} & \textcolor{black}{71.6} & \textcolor{black}{36.4} \\
\rowcolor[HTML]{EFEFEF} 
4 & \textcolor{black}{121.3} & \textcolor{black}{84.7} & \textcolor{black}{112.4} & \textcolor{black}{43.7} \\
Growth $M{=}4/1$ & \textcolor{black}{$5.0\times$} & \textcolor{black}{$4.3\times$} & \textcolor{black}{$4.4\times$} & \textcolor{black}{$2.5\times$} \\
\midrule
Miss\% at $M{=}4$ & \textcolor{black}{61.8} & \textcolor{black}{42.3} & \textcolor{black}{53.7} & \textcolor{black}{0.0} \\
\bottomrule
\Xhline{1.2pt}
\end{tabular}
\end{adjustbox}
\end{table}

\textcolor{black}{Table~\ref{tab:xrf_scaling} confirms that PAIP's scaling property transfers to XRF55. With occupancy $M$ taken from the Kinect skeleton stream and motion support recovered from the released RD map, PRISM grows from \textcolor{black}{17.8}~ms at $M{=}1$ to \textcolor{black}{43.7}~ms at $M{=}4$, a \textcolor{black}{$2.5\times$} increase for a fourfold change in occupancy, and maintains a \textcolor{black}{0.0}\% miss rate throughout. The baselines again grow superlinearly and violate the deadline at $M{=}4$, reproducing the ordering observed on RT-Pose. This dataset also lets us exercise the truncation path of PAIP Step~4 directly: when the active profile is Balanced with $M_{\max}{=}3$ and a frame contains \textcolor{black}{$M{=}4$} persons, PAIP retains the three highest-energy proposals and the measured latency remains bounded at \textcolor{black}{34.1}~ms with \textcolor{black}{0.0}\% misses, at the cost of a recall reduction to \textcolor{black}{0.78}. The truncation therefore preserves the timing guarantee exactly as designed while making its completeness cost explicit and measurable.}

\begin{table}[t]
\caption{\textcolor{black}{DAOP deadline-aware selection on mmRadPose (single-person) under five deadline budgets. PRISM-ND uses a fixed Balanced profile. }}
\label{tab:daop_cross}
\centering
\begin{adjustbox}{width=\linewidth,center}
\setlength{\tabcolsep}{1pt}
\begin{tabular}{c|ccccc}
\Xhline{1.2pt}
\toprule
$T_d$ (ms) & \textbf{DAOP} & $T_{\text{WCET}}^{\text{pred}}$ & \textbf{MPJPE} (mm) & \textbf{PRISM Miss\%} & \textbf{ND Miss\%} \\
\midrule
55 & \textcolor{black}{Ultra-Precise} & \textcolor{black}{54.8} & \textcolor{black}{46.2} & \textcolor{black}{0.0} & \textcolor{black}{0.0} \\
\rowcolor[HTML]{EFEFEF} 
45 & \textcolor{black}{Precise}       & \textcolor{black}{44.2} & \textcolor{black}{49.8} & \textcolor{black}{0.0} & \textcolor{black}{0.0} \\
35 & \textcolor{black}{Balanced}      & \textcolor{black}{33.6} & \textcolor{black}{57.1} & \textcolor{black}{0.0} & \textcolor{black}{0.0} \\
\rowcolor[HTML]{EFEFEF} 
23 & \textcolor{black}{Light}         & \textcolor{black}{22.4} & \textcolor{black}{74.3} & \textcolor{black}{0.0} & \textcolor{black}{19.4} \\
14 & \textcolor{black}{Ultra-Light}   & \textcolor{black}{13.8} & \textcolor{black}{91.6} & \textcolor{black}{0.0} & \textcolor{black}{73.1} \\
\bottomrule
\Xhline{1.2pt}
\end{tabular}
\end{adjustbox}
\end{table}

\textcolor{black}{Table~\ref{tab:daop_cross} confirms that DAOP's deadline-aware behavior likewise transfers to mmRadPose under the same unified RAD lattice. Across all five deadline budgets DAOP selects the highest-accuracy feasible profile and sustains a \textcolor{black}{0.0}\% miss rate, whereas the fixed-profile PRISM-ND begins to violate deadlines once the budget falls below its certified Balanced cost, reaching \textcolor{black}{19.4}\% misses at $T_d{=}23$~ms and \textcolor{black}{73.1}\% at $T_d{=}14$~ms. The accuracy recovered under loose budgets is also consistent with the RT-Pose result, since DAOP attains \textcolor{black}{46.2}~mm at $T_d{=}55$~ms against the \textcolor{black}{57.1}~mm of the fixed Balanced profile. Taken together, these two experiments indicate that the scaling behavior of PAIP and the deadline-aware control of DAOP reproduce across front-ends and ground-truth pipelines.}

\section{Real-World Prototype Deployment}
\label{sec:prototype}

The evaluation in Section~\ref{sec:eval} streams public datasets from local storage to ensure a controlled and reproducible comparison across all systems. To verify that PRISM operates end-to-end on a physical sensor rather than only on pre-recorded data, we additionally deploy it in live-capture mode on a self-assembled edge prototype and measure its behavior on radar frames acquired in real time. This section reports the prototype configuration, the live timing and accuracy results, and a failure-case analysis.

\subsection{Prototype Configuration}
\label{sec:proto_setup}

The prototype pairs the same Raspberry Pi~5 used throughout Section~\ref{sec:eval} with a Texas Instruments AWR1843BOOST FMCW mmWave radar operating in the 76--81~GHz band. The radar streams raw frames to the Raspberry Pi over USB; frame-arrival timestamps are captured at the application layer before any computation begins, so that DMA and USB transfer latencies are included in the reported end-to-end timing. The radar is configured to produce a Range-Angle-Doppler tensor compatible with the PRISM front-end, and the operating profiles are recalibrated for this platform following the offline procedure of Section~\ref{sec:design_profiles} (the platform constants $(c_1,c_2,c_3)$ are re-derived from worst-case micro-benchmarks on this radar-Raspberry Pi pairing). For quantitative accuracy reference only, an RGB-D camera is co-located with the radar to provide synchronized skeletal annotations; this camera is used solely to obtain reference joint positions for evaluation and is not part of the deployed inference pipeline, which consumes radar input exclusively.

\begin{figure*}[t]
  \centering
  \includegraphics[width=\linewidth]{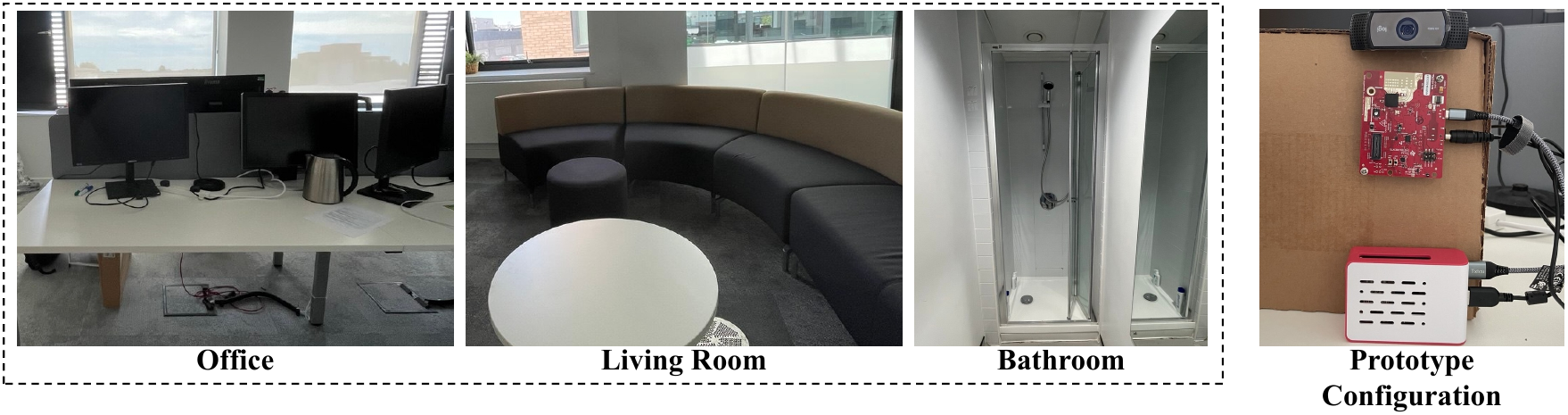}
  \caption{Real-world prototype deployment across three indoor environments. The edge node pairs a Raspberry Pi~5 with a TI AWR1843BOOST mmWave radar; an RGB-D camera is co-located to provide reference annotations and is not part of the radar-only inference pipeline. (a)~Office, (b)~Living Room, (c)~Bathroom.}
  \label{fig:proto_scenarios}
\end{figure*}

We collect short sequences across three indoor environments with distinct geometry and clutter, namely a bathroom, a living room, and an office building, as illustrated in Figure~\ref{fig:proto_scenarios}. These environments differ in room size, furniture density, and multipath conditions, and therefore stress the cross-scene stability of the Doppler-based gating (O4) under genuine sensor noise rather than dataset-curated frames. Data collection was approved by our institution's research ethics committee; all participants were lab members who gave informed consent, and the co-located RGB-D footage was used only offline to compute reference joint positions and is not distributed outside the research team.

\subsection{Live-Capture Timing and Accuracy}
\label{sec:proto_results}

Table~\ref{tab:prototype} reports PRISM's live-capture performance against the dataset-replay reference, with DAOP selecting the Precise profile at $T_d{=}45$~ms. The live MPJPE is computed against the RGB-D reference skeletons across the three environments.

\begin{table}[t]
\caption{Live-capture prototype results (Raspberry Pi~5 + AWR1843BOOST), $T_d{=}45$~ms, DAOP selects Precise. Dataset-replay reference reproduced from Table~\ref{tab:e2e}. }
\label{tab:prototype}
\centering
\begin{adjustbox}{width=\linewidth,center}
\setlength{\tabcolsep}{1pt}
\begin{tabular}{l|ccccc}
\Xhline{1.2pt}
\toprule
\textbf{Setting} & \textbf{p99(ms)} & \textbf{Max(ms)} & \textbf{Miss\%} & \textbf{Mem(MB)} & \textbf{MPJPE(mm)} \\
\midrule
Dataset-replay & \textcolor{black}{37.4} & \textcolor{black}{39.2} & \textbf{0.0} & \textcolor{black}{298} & \textcolor{black}{55.2} \\
\rowcolor[HTML]{EFEFEF} 
Live: Office    & \textcolor{black}{39.6} & \textcolor{black}{42.8} & \textcolor{black}{0.0} & \textcolor{black}{305} & \textcolor{black}{61.4} \\
Live: Living Room      & \textcolor{black}{40.2} & \textcolor{black}{43.5} & \textcolor{black}{0.0} & \textcolor{black}{307} & \textcolor{black}{64.7} \\
\rowcolor[HTML]{EFEFEF} 
Live: Bathroom       & \textcolor{black}{41.1} & \textcolor{black}{44.3} & \textcolor{black}{0.2} & \textcolor{black}{309} & \textcolor{black}{66.3} \\
\bottomrule
\Xhline{1.2pt}
\end{tabular}
\end{adjustbox}
\end{table}

Across all three live environments, PRISM keeps p99 latency within the Precise-profile WCET margin while including DMA and USB acquisition overhead. Office and living-room recordings record a \textcolor{black}{0.0}\% miss rate; the bathroom recording records \textcolor{black}{0.2}\% misses, discussed below. The observed p99 latency increases by \textcolor{black}{2.2}--\textcolor{black}{3.7}~ms relative to the dataset-replay reference, consistent with the frame-acquisition and transfer cost that is absent when frames are read from local storage. The live MPJPE is \textcolor{black}{6.2}--\textcolor{black}{11.1}~mm higher than the dataset-replay reference, attributable to real-sensor noise, radar--camera spatial calibration, and the angular resolution of the AWR1843BOOST; the live estimates remain more accurate than the deadline-missing baselines in Table~\ref{tab:e2e}.




The live deployment exposes failure modes that are under-represented in curated datasets. We observe three recurring categories. 

\emph{First, near-stationary targets.} When a subject remains nearly motionless and faces the radar directly, the Doppler gate (PAIP Step~1) attenuates the weak reflection and may fail to generate a proposal, consistent with the limitation discussed in Section~\ref{app:limitations}; in the live sequences this accounts for the majority of missed detections, and brief stillness gaps are partially bridged by carrying proposals forward from adjacent active frames. 
\emph{Second, strong multipath in confined geometry.} In the bathroom environment, compact layout and reflective surfaces produce specular multipath that occasionally generates a spurious dynamic component passing the Doppler gate; the anthropometric-area filter (PAIP Step~3) rejects most such artifacts, but residual cases inflate per-frame cost and account for the \textcolor{black}{0.2}\% deadline miss reported in Table~\ref{tab:prototype}. 
\emph{Third, close-proximity merging.} When two subjects stand within approximately one body width, their motion energy merges into a single connected component, and PRISM produces one skeleton for the merged region rather than two, consistent with the merge-rate analysis in Section~\ref{sec:eval_paip}. 

None of these failure modes moves live p99 outside the Precise-profile WCET margin reported above; they primarily affect detection completeness or per-frame accuracy, which is consistent with PRISM's priority of preserving a bounded deadline under live input.

\section{Additional Robustness Studies}
\label{app:robustness}

\subsection{PRISM-SWG: Sparse-ROI Sliding-Window Ablation}
\label{sec:eval_swg}

PRISM-SWG isolates the contribution of PBIP's integral queries from the general benefit of physics-aware ROI reduction. It retains PAIP's Doppler-gated proposal and spatial masking but replaces the constant-count summed-volume lookup with a conventional sliding-window operator applied \emph{within the masked ROI only}. This ablation directly tests whether the constant-time query property of PBIP matters once the workset has already been reduced by PAIP.


\begin{figure}[t]
  \centering
  \includegraphics[width=\columnwidth]{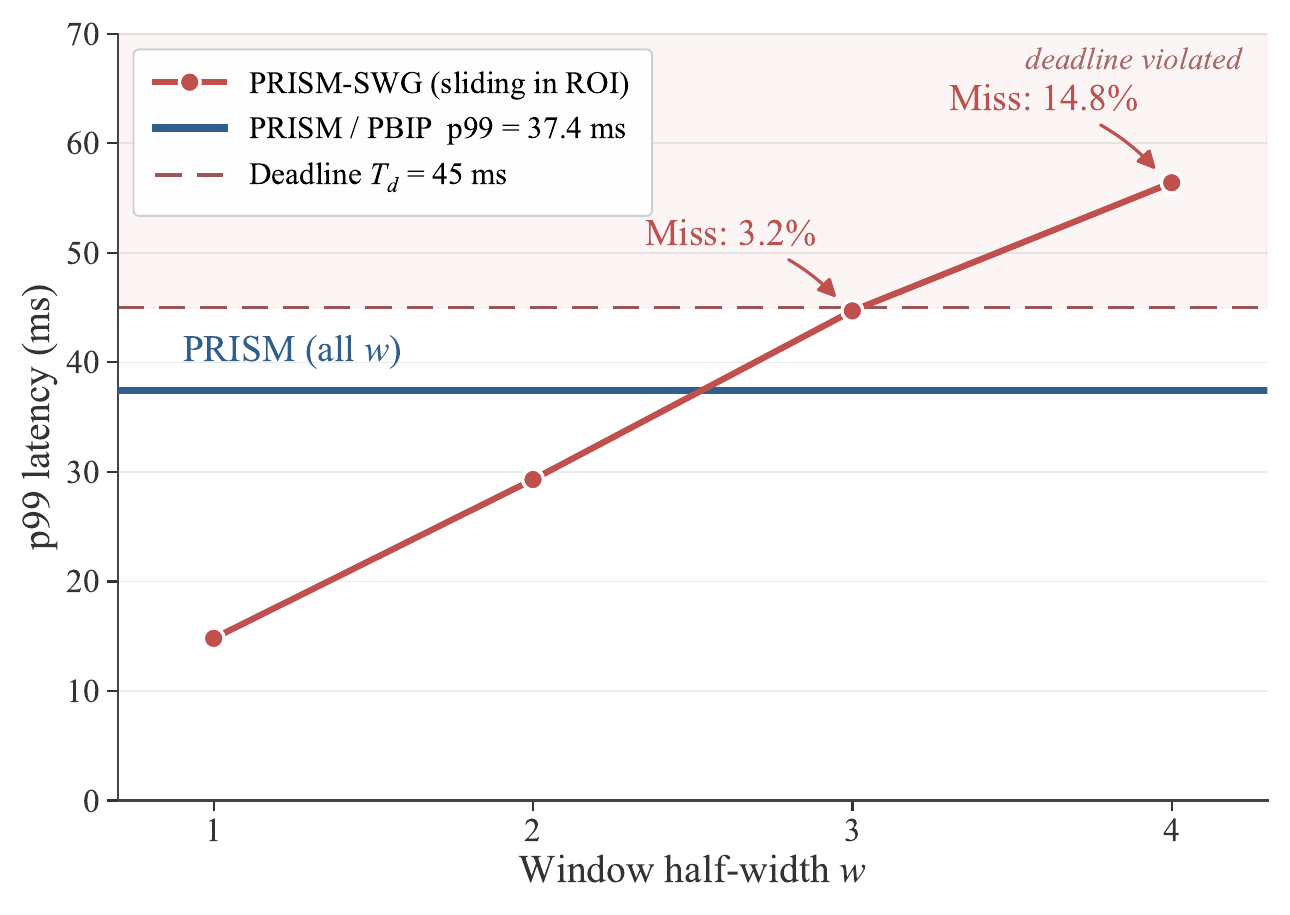}
  \caption{PRISM-SWG vs.\ PRISM on the \rtposeedit{RT-Pose multi-person split}, $M{=}2$, $T_d{=}45$~ms. SWG applies a sliding window within the PAIP-masked ROI; PRISM uses PBIP integral queries.}
  \label{fig:swg}
\end{figure}

Figure~\ref{fig:swg} shows that PRISM-SWG at $w{=}1$ meets the deadline (\rtposeedit{0.0}\% misses) because the masked ROI is small enough for a $3{\times}3$ window to complete within budget. However, at $w{\ge}3$ the $(2w{+}1)^2$ scaling within the ROI drives latency beyond $T_d$, accumulating \rtposeedit{3.2}\% misses at $w{=}3$ and \rtposeedit{14.8}\% at $w{=}4$. PRISM with PBIP achieves comparable accuracy (\rtposeedit{55.2}~mm vs.\ \rtposeedit{53.6}~mm at $w{=}4$) while remaining strictly within deadline at all window sizes, confirming that the integral-query property, rather than spatial ROI reduction alone, is the structural source of window-independent timing.

\subsection{EVT Tail Fitting Validation}
\label{sec:eval_evt}

The WCET construction in Eq.~\eqref{eq:WCET} uses a fixed 5\% margin atop the 99.9th percentile micro-benchmark. To assess whether formal extreme-value theory (EVT) would yield materially different bounds, we fit Generalized Pareto Distribution (GPD) models to the upper tail of execution time samples for each profile (\rtposeedit{10,000} frames each) and extract the $10^{-6}$ exceedance quantile.

\begin{table}[t]
\caption{EVT-derived bounds vs.\ fixed margin bounds on the \rtposeedit{RT-Pose multi-person split}. GPD fitted to the upper 5\% tail of execution times for each profile. $T_{\text{EVT}}^{10^{-6}}$: GPD quantile at $10^{-6}$ exceedance probability.}
\label{tab:evt}
\centering
\begin{adjustbox}{width=\linewidth,center}
\setlength{\tabcolsep}{7pt}
\begin{tabular}{l|cccc}
\Xhline{1.2pt}
\toprule
\textbf{Profile} & $T_{\text{WCET}}^{\text{pred}}$ & $T_{\text{EVT}}^{10^{-6}}$ & $\Delta$ (ms) & \textbf{Shape $\xi$} \\
\midrule
Ultra-Light   & \textcolor{black}{13.8} & \textcolor{black}{14.2} & \textcolor{black}{$+$0.4} & \textcolor{black}{$-$0.12} \\
\rowcolor[HTML]{EFEFEF} 
Light         & \textcolor{black}{22.4} & \textcolor{black}{23.1} & \textcolor{black}{$+$0.7} & \textcolor{black}{$-$0.09} \\
Balanced      & \textcolor{black}{33.6} & \textcolor{black}{34.4} & \textcolor{black}{$+$0.8} & \textcolor{black}{$-$0.11} \\
\rowcolor[HTML]{EFEFEF} 
Precise       & \textcolor{black}{44.2} & \textcolor{black}{45.3} & \textcolor{black}{$+$1.1} & \textcolor{black}{$-$0.08} \\
Ultra-Precise & \textcolor{black}{54.8} & \textcolor{black}{56.1} & \textcolor{black}{$+$1.3} & \textcolor{black}{$-$0.10} \\
\bottomrule
\Xhline{1.2pt}
\end{tabular}
\end{adjustbox}
\end{table}

Table~\ref{tab:evt} shows that the EVT-derived $10^{-6}$ bounds exceed the fixed-margin predictions $T_{\text{WCET}}^{\text{pred}}$ by only \rtposeedit{0.4}--\rtposeedit{1.3}~ms, and all GPD shape parameters $\xi$ are negative (\rtposeedit{$-$0.12} to \rtposeedit{$-$0.08}), indicating a bounded upper tail rather than a heavy-tailed distribution. This confirms that the fixed 5\% margin is conservative \emph{relative to the 99.9th-percentile measurement it is built from}, and that the execution time distribution is consistent with the bounded loop structure of PBIP and PAIP.

\textcolor{black}{A stricter and more informative test is whether $T_{\text{EVT}}^{10^{-6}}$ also stays below the deadline $T_d$ actually assigned to each profile in Table~\ref{tab:daop_select}. It does not, in four of the five settings: Ultra-Light (\rtposeedit{14.2} vs.\ $T_d{=}14$~ms), Light (\rtposeedit{23.1} vs.\ $T_d{=}23$~ms), Precise (\rtposeedit{45.3} vs.\ $T_d{=}45$~ms), and Ultra-Precise (\rtposeedit{56.1} vs.\ $T_d{=}55$~ms) all exceed their assigned deadline at the $10^{-6}$ exceedance quantile, and only Balanced remains strictly below its deadline (\rtposeedit{34.4} vs.\ $T_d{=}35$~ms). This does not contradict the \rtposeedit{0.0}\% empirical miss rate reported over the evaluated \rtposeedit{10,000}-frame traces in Table~\ref{tab:daop_select}, since no sample in those traces reached the $10^{-6}$ tail; rather, it shows that the fixed 5\% margin should not be read as a deterministic certificate against $T_d$ itself. We therefore state PRISM's timing guarantee precisely: consistent with the MBPTA paradigm it follows~\cite{cucu2012measurement}, $T_{\text{WCET}}^{(p)}$ is a measurement-based, high-confidence probabilistic timing bound, and the deadline compliance reported throughout this paper is an empirical property validated over the evaluated traces rather than a proof that exceedance at arbitrarily small probability is impossible. Closing this gap so that $T_{\text{EVT}}^{10^{-6}} \le T_d$ holds uniformly, for instance by widening the fixed margin from 5\% to approximately 8\%, is a direction we adopt for deployments that require a formally certified rather than probabilistically bounded guarantee.}

\subsection{Co-Runner Interference Robustness}
\label{sec:eval_corunner}

The WCET bounds in Table~\ref{tab:profile_wcet} were derived under isolated single-task execution. In realistic edge deployments, PRISM may share CPU resources with other tasks. We therefore evaluate timing robustness under controlled co-runner contention by running \texttt{stress-ng} memory and cache stressors on 3 of the 4 Cortex-A76 cores while PRISM executes on the remaining core with CPU affinity (\texttt{taskset}) on the \rtposeedit{RT-Pose multi-person split}, $M{=}2$, $T_d{=}45$~ms.

\begin{table}[t]
\caption{Timing impact of co-runner interference. PRISM (Precise) on 1 core, \texttt{stress-ng} on 3 cores, \rtposeedit{RT-Pose multi-person split}, $M{=}2$, $T_d{=}45$~ms.}
\label{tab:corunner}
\centering
\begin{adjustbox}{width=\linewidth,center}
\setlength{\tabcolsep}{1pt}
\begin{tabular}{l|cccc}
\Xhline{1.2pt}
\toprule
\textbf{Condition} & \textbf{p99 (ms)} & \textbf{Max (ms)} & \textbf{Miss\%} & \textbf{MPJPE (mm)} \\
\midrule
Isolated          & \textcolor{black}{37.4} & \textcolor{black}{39.2} & \textbf{0.0} & \textcolor{black}{55.2} \\
\rowcolor[HTML]{EFEFEF} 
Co-runner (mem)   & \textcolor{black}{39.8}  & \textcolor{black}{42.6}  & \textbf{0.0}  & \textcolor{black}{55.2} \\
Co-runner (cache) & \textcolor{black}{41.3}  & \textcolor{black}{44.1}  & \textbf{0.0}  & \textcolor{black}{55.2} \\
\rowcolor[HTML]{EFEFEF} 
Co-runner (both)  & \textcolor{black}{42.7}  & \textcolor{black}{46.8}  & \textcolor{red!70!black}{2.1}  & \textcolor{black}{55.2} \\
\bottomrule
\Xhline{1.2pt}
\end{tabular}
\end{adjustbox}
\end{table}

Table~\ref{tab:corunner} shows that under memory-only or cache-only contention, PRISM maintains \rtposeedit{0.0}\% deadline misses with p99 increases of \rtposeedit{2.4}--\rtposeedit{3.9}~ms attributable to shared L3 cache evictions and memory bus contention. Under combined memory-plus-cache stress, the maximum observed latency reaches \rtposeedit{46.8}~ms, producing \rtposeedit{2.1}\% misses. This result indicates that the isolated-run WCET table is not sufficient under unmanaged co-runner contention. Deployments that require the same guarantee under interference should recalibrate the safety margin under the target workload mix or reserve CPU and memory bandwidth explicitly. Accuracy is unaffected (\rtposeedit{55.2}~mm in all conditions) because co-runner contention delays computation but does not alter the bounded workset or model weights.

\subsection{Case Study: Continuous Operation Under Occupancy}
\label{sec:case_study}

Held-out tables on the \rtposeedit{RT-Pose multi-person split}, including the per-deadline sweep in Table~\ref{tab:daop_select}, score frames independently and therefore do not show a long mixed-occupancy recording. We replay a continuous 30-minute \rtposeedit{RT-Pose} segment comprising \rtposeedit{27,000} frames at 15~fps, in which person count varies between 1 and 3, under a fixed deadline $T_d{=}35$~ms.

Over this trace, PRISM records zero deadline misses. HuPRModel and milliMamba produce \rtposeedit{4,832} and \rtposeedit{3,147} misses, concentrated at $M{\ge}3$. PRISM-ND, which keeps the Balanced profile throughout, accumulates \rtposeedit{2,416} misses (\rtposeedit{8.9}\%) on the same recording. The comparison with HuPRModel and milliMamba indicates that physics-bounded per-instance processing keeps tail latency inside the deadline under occupancy change. Table~\ref{tab:daop_select} reports \rtposeedit{0.0}\% ND misses at the same $T_d{=}35$~ms on independently scored split frames; the contiguous stream additionally includes occupancy-transition frames those tables do not isolate, so a frozen Balanced profile has less slack on this recording while the full pipeline remains inside the deadline. The per-deadline selection rule is unchanged from Table~\ref{tab:daop_select}.

\begin{figure}[t]
  \centering
  \includegraphics[width=\columnwidth]{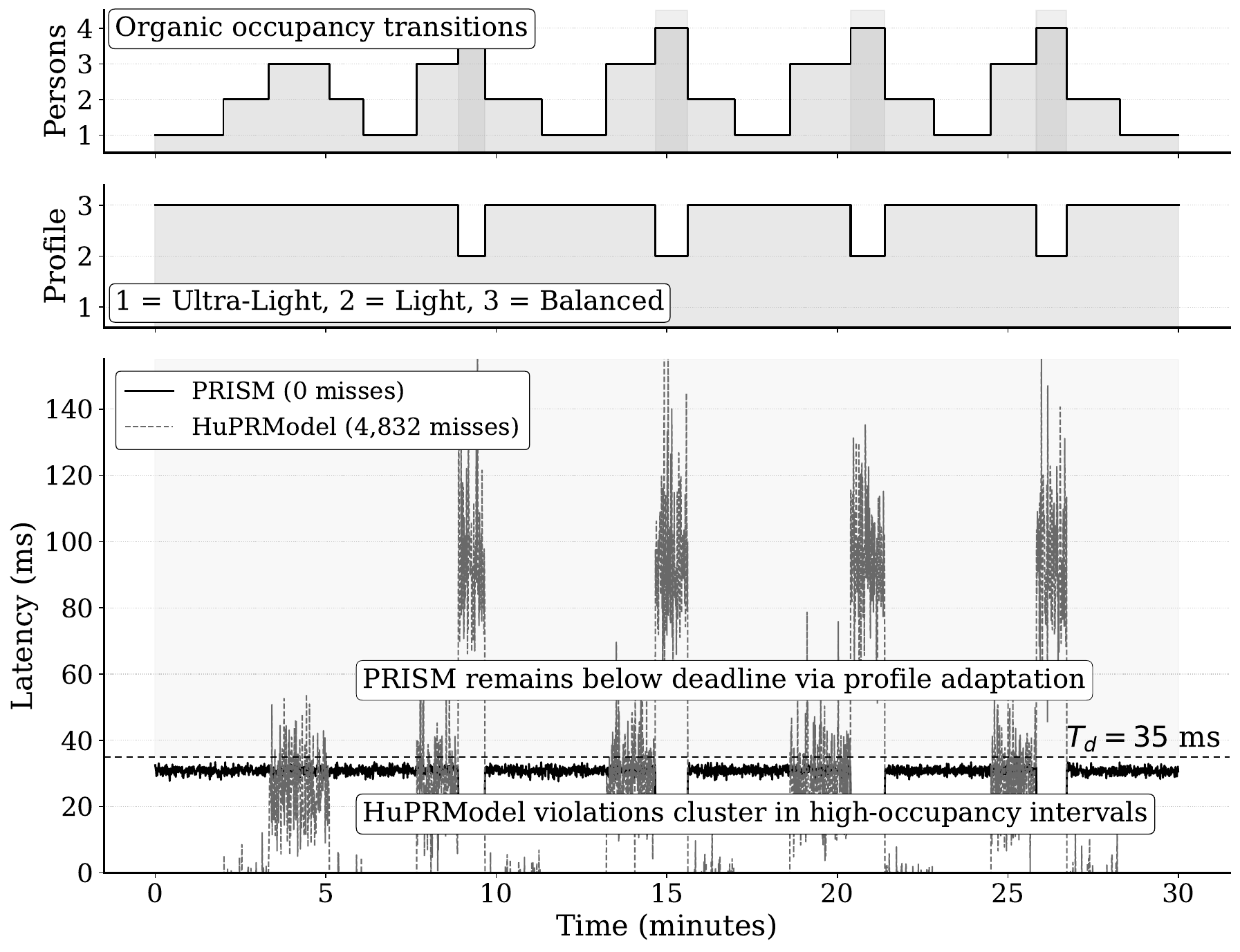}
  \caption{Case study on a continuous \rtposeedit{RT-Pose} segment ($T_d{=}35$~ms). Top: person count over time. Middle: selected profile index under the fixed deadline. Bottom: execution time for each frame for PRISM (solid) and HuPRModel (dashed) vs.\ $T_d$ (horizontal). PRISM records zero misses.}
  \label{fig:case}
\end{figure}

Zero misses across the full recording are consistent with the coordinated PBIP--PAIP--DAOP stack under a fixed deadline: PBIP makes per-stage cost certifiable, PAIP bounds per-instance workload, and DAOP supplies the deadline-feasible profile used for the trace. Each component remains necessary for this combined behavior.

\section{Accuracy Degradation Under Multi-Person Load}
\label{app:accuracy_degradation}

Table~\ref{tab:accuracy_vs_M} complements the timing analysis with MPJPE across $M{=}1$--$3$ on the \rtposeedit{RT-Pose multi-person split}. The results confirm that accuracy and timing degradation share common architectural roots.

\begin{table}[t]
\caption{MPJPE (mm) vs.\ active person count $M$ on the \rtposeedit{RT-Pose multi-person split}. Single-person ($M{=}1$) values serve as the degradation baseline. Lower is better; $\Delta(M{=}3{-}M{=}1)$ and ratio are relative to $M{=}1$.}
\label{tab:accuracy_vs_M}
\centering
\begin{adjustbox}{width=\linewidth,center}
\setlength{\tabcolsep}{2pt}
\begin{tabular}{l|cccccc}
\Xhline{1.2pt}
\toprule
\textbf{Method} & $M{=}1$ & $M{=}2$ & $M{=}3$ & $\Delta(M{=}3{-}M{=}1)$ & Ratio \\
\midrule
HuPRModel  & \textcolor{black}{78.4}  & \textcolor{black}{89.3}  & \textcolor{black}{112.7} & \textcolor{black}{$+$34.3} & 
\textcolor{black}{$\times$1.44} \\
\rowcolor[HTML]{EFEFEF} 
RETR       & \textcolor{black}{78.1}  & \textcolor{black}{86.4}  & \textcolor{black}{103.8} & \textcolor{black}{$+$25.7} & \textcolor{black}{$\times$1.33} \\
mmDiff     & \textcolor{black}{75.5}  & \textcolor{black}{82.1}  & \textcolor{black}{98.6}  & \textcolor{black}{$+$23.1} & \textcolor{black}{$\times$1.31} \\
\rowcolor[HTML]{EFEFEF} 
MVDoppler & \textcolor{black}{75.6} & \textcolor{black}{81.8} & \textcolor{black}{91.4} & \textcolor{black}{$+$15.8} & \textcolor{black}{$\times$1.21} \\
milliMamba & \textcolor{black}{76.3}  & \textcolor{black}{84.7}  & \textcolor{black}{108.3} & \textcolor{black}{$+$32.0} & \textcolor{black}{$\times$1.42} \\
\rowcolor[HTML]{EFEFEF} 
SMH        & \textcolor{black}{64.9}  & \textcolor{black}{83.4}  & \textcolor{black}{107.6}  & \textcolor{black}{$+$42.7}  & \textcolor{black}{$\times$1.66}  \\
\midrule
PRISM-NB   & \textcolor{black}{50.2}  & \textcolor{black}{54.8}  & \textcolor{black}{61.4}  & \textcolor{black}{$+$11.2} & \textcolor{black}{$\times$1.22} \\
\textbf{PRISM (Precise)} & \textbf{\textcolor{black}{53.1}} & \textbf{\textcolor{black}{55.2}} & \textbf{\textcolor{black}{57.8}} & \textbf{\textcolor{black}{$+$4.7}} & \textbf{\textcolor{black}{$\times$1.09}} \\
\bottomrule
\Xhline{1.2pt}
\end{tabular}
\end{adjustbox}
\end{table}

Methods without per-instance spatial gating suffer the steepest declines even within the observed $M{=}1$--$3$ range: \textbf{HuPRModel} ($\times$\textcolor{black}{1.44}) conflates multi-body activations in a shared global encoder; \textbf{RETR} ($\times$\textcolor{black}{1.33}) exhibits the same timing-accuracy coupling via $O(n^2)$ cross-attention spanning all persons simultaneously; and \textbf{mmDiff} ($\times$\textcolor{black}{1.31}) denoises from an increasingly multi-modal scene posterior whose fixed step budget cannot resolve superimposed hypotheses. Sequential models are equally vulnerable: \textbf{milliMamba} ($\times$\textcolor{black}{1.42}) propagates trajectory estimates calibrated to single-person dynamics, so entry/exit events produce persistent state discontinuities; \textbf{SMH} ($\times$\textcolor{black}{1.66}), despite its accuracy-efficient physics-guided pipeline, collapses under multi-person conditions because its SSP/MCP/HMSF modules aggregate scene-level energy into a single fixed-size feature vector without per-instance isolation, leaving the MLP regressor with a blended multi-person input it cannot decompose. \textbf{MVDoppler} shows the mildest baseline degradation ($\times$\textcolor{black}{1.21}) through partial motion isolation, but inter-sensor angular diversity still breaks down under mutual occlusion, consistent with its \textcolor{black}{14.2}\% deadline miss rate at $M{=}3$.

PRISM achieves a $\times$\textcolor{black}{1.09} ratio ($+$\textcolor{black}{4.7}~mm), versus the best-baseline \textcolor{black}{1.21}. This near-invariance stems from PBIP's per-person physics-derived workset $\mathcal{V}_{\text{body}}^{(i)}$, which isolates feature extraction at the signal level independently for each instance. The ablation comparison with PRISM-NB ($\times$\textcolor{black}{1.22}) confirms the attribution: retaining PAIP's scheduling isolation but substituting PBIP's integral queries with unconstrained aggregation causes accuracy degradation consistent with the heavyweight baselines.

\section{Stability Across Scenes}
\label{app:cross_scene}

\begin{table}[t]
\caption{Per-scene timing and accuracy variation across the \rtposeedit{RT-Pose multi-person split} ($M{=}2$, $T_d{=}45$~ms). $\Delta_T$: maximum absolute deviation of per-scene mean latency from $\mu_T$ (ms); $\Delta_E$: maximum absolute deviation of per-scene MPJPE from $\mu_E$ (mm).}
\label{tab:cross_scene}
\centering
\begin{adjustbox}{width=\linewidth,center}
\setlength{\tabcolsep}{4pt}
\begin{tabular}{l|cccc}
\Xhline{1.2pt}
\toprule
\textbf{Method} & $\mu_T$ (ms) & $\Delta_T$ (ms) & $\mu_E$ (mm) & $\Delta_E$ (mm) \\
\midrule
HuPRModel      & \textcolor{black}{39.4} & \textcolor{black}{$\pm$12.3} & \textcolor{black}{89.3} & \textcolor{black}{$\pm$9.2} \\
\rowcolor[HTML]{EFEFEF} 
milliMamba     & \textcolor{black}{33.6} & \textcolor{black}{$\pm$8.7}  & \textcolor{black}{84.7} & \textcolor{black}{$\pm$7.1} \\
MVDoppler & \textcolor{black}{32.1} & \textcolor{black}{$\pm$6.4}  & \textcolor{black}{81.8} & \textcolor{black}{$\pm$5.8} \\
\rowcolor[HTML]{EFEFEF} 
\textbf{PRISM (Precise)} & \textbf{\textcolor{black}{27.3}} & \textbf{\textcolor{black}{$\pm$1.6}} & \textbf{\textcolor{black}{55.2}} & \textbf{\textcolor{black}{$\pm$3.8}} \\
\bottomrule
\Xhline{1.2pt}
\end{tabular}
\end{adjustbox}
\end{table}

Table~\ref{tab:cross_scene} reports timing and accuracy variation across scenes on the \rtposeedit{RT-Pose multi-person split}. PRISM (Precise) latency varies by at most \rtposeedit{$\pm$1.6}~ms and MPJPE by at most \rtposeedit{$\pm$3.8}~mm. This stability follows from PAIP's Doppler-based motion gating (Section~\ref{sec:design_paip}), which suppresses static background regardless of scene geometry and makes the active map insensitive to clutter distribution in each scene. HuPRModel exhibits \rtposeedit{$\pm$12.3}~ms latency variation, reflecting uncontrolled sensitivity to point density in each scene. This robustness across scenes directly validates O4 (Section~\ref{sec:observations}): the Doppler threshold transfers across sites without recalibration for each scene.

\section{Generalization Across Datasets}
\label{app:cross_dataset}

To validate that PRISM's physics-bounded guarantees are not artifacts of a specific radar configuration, we evaluate on two additional datasets with contrasting signal characteristics.

\textbf{XRF55 (front-end generalization).} XRF55 releases RA and RD maps from which a RAD tensor is reconstructed, as described in Section~\ref{sec:eval_setup}. Table~\ref{tab:cross} reports MPJPE under this construction. PBIP's integral queries operate on the reconstructed support and retain their $O(1)$ cost; occupancy $M$ follows the synchronized Kinect skeleton count. The accuracy difference relative to HuPR is modest (\textcolor{black}{$+$5.3}~mm), while several baselines that process a dense tensor still miss the deadline.

\textbf{mmRadPose (OMC supervision).} mmRadPose provides optical motion capture ground truth for 12~subjects~\cite{mueller2025radproposer}. The input is the same explicit RAD tensor used for HuPR, obtained by standard FMCW processing of the released radar cubes. Table~\ref{tab:cross} reports MPJPE on mmRadPose. PRISM achieves \textcolor{black}{49.8}~mm MPJPE while maintaining \textcolor{black}{0.0}\% deadline misses, indicating that both accuracy and schedulability transfer under high-precision motion-capture supervision.

\begin{table}[t]
\caption{Generalization across datasets. MPJPE (mm) on XRF55 and mmRadPose. $\dagger$ = schedulability violation ($>$0\% miss rate). }
\label{tab:cross}
\centering
\begin{adjustbox}{width=\linewidth,center}
\setlength{\tabcolsep}{8pt}
\begin{tabular}{l|cccc}
\Xhline{1.2pt}
\toprule
\textbf{Method} & \multicolumn{2}{c}{\textbf{XRF55}} & \multicolumn{2}{c}{\textbf{mmRadPose}} \\
 & \textbf{MPJPE} & \textbf{Miss\%} & \textbf{MPJPE} & \textbf{Miss\%} \\
\midrule
HuPRModel$^\dagger$ & \textcolor{black}{84.7} & \textcolor{black}{8.4} & \textcolor{black}{81.2} & \textcolor{black}{6.3} \\
\rowcolor[HTML]{EFEFEF} 
RETR$^\dagger$ & \textcolor{black}{81.6} & \textcolor{black}{6.7} & \textcolor{black}{79.3} & \textcolor{black}{5.1} \\
milliMamba$^\dagger$ & \textcolor{black}{86.4} & \textcolor{black}{3.8} & \textcolor{black}{77.8} & \textcolor{black}{3.4} \\
\rowcolor[HTML]{EFEFEF} 
SMH$^\dagger$     & \textcolor{black}{73.1} & \textcolor{black}{2.3} & \textcolor{black}{62.7} & \textcolor{black}{1.8} \\
\textbf{PRISM (Precise)} & \textbf{\textcolor{black}{58.4}} & \textbf{\textcolor{black}{0.0}} & \textbf{\textcolor{black}{49.8}} & \textbf{\textcolor{black}{0.0}} \\
\bottomrule
\Xhline{1.2pt}
\end{tabular}
\end{adjustbox}
\end{table}

\section{Baseline Adaptation and Comparison Fairness}
\label{app:baseline_fairness}

The methods compared in Section~\ref{sec:eval} originate from heterogeneous design assumptions about input representation, sensor configuration, and target task. A direct transplantation of their released implementations would therefore conflate genuine algorithmic differences with incidental mismatches in input format or hardware budget. To ensure that the reported timing and accuracy differences reflect the algorithms themselves rather than such mismatches, we adopt a single adaptation protocol governed by three principles. First, every method consumes the same per-dataset input representation, so that no baseline receives a privileged signal. Second, every method is retrained from scratch on each dataset under its official optimization procedure, so that no comparison relies on weights transferred from a different sensor or task. Third, every method executes in the identical single-threaded CPU environment described in Section~\ref{sec:eval_setup}, so that latency reflects an equal hardware budget. The remainder of this section specifies how each principle is realized and why the resulting comparison is fair.

\subsection{Unified Input Representation}

All evaluated methods operate on the Range-Angle-Doppler tensor defined in Section~\ref{sec:primer}, constructed identically for every method on a given dataset. For XRF55 the shared tensor is reconstructed from the released RA and RD maps; for HuPR and mmRadPose it is obtained by standard FMCW processing of the released radar cubes~\cite{lee2023hupr,wang2024xrf55,mueller2025radproposer}. Methods whose original pipelines consume a CFAR point cloud rather than a dense tensor, such as mmDiff, are supplied with a point cloud derived from this same tensor through the standard detection stage, so that the information available to them remains a deterministic function of the shared input rather than an independently curated representation. Methods whose original pipelines assume multiple synchronized radar views, namely HuPRModel, RETR, and MVDoppler, have their multi-view aggregation modules removed and their single-view processing paths retained. This adaptation preserves each method's core modeling mechanism while removing a sensing requirement that the single-sensor deployment scenario of this work does not provide. Because the removed components concern only the fusion of additional viewpoints, the retained pipeline remains faithful to the original feature-extraction intent.

RT-Pose requires an additional alignment step, since it is released as a Cartesian four-dimensional tensor rather than a polar RAD tensor. We remap it once into the RAD-aligned representation with an explicit Doppler axis described in Section~\ref{sec:eval_setup}, and this single remapped representation is then provided to every method without exception. Consequently, any residual information loss introduced by the remapping is borne equally by all methods, and the relative comparison on RT-Pose remains internally consistent. PRISM derives no advantage that is unavailable to the baselines, because the Doppler-resolved structure required by PAIP and PBIP is exposed in exactly the same tensor that the baselines receive.

\subsection{Per-Baseline Adaptation Rationale}

Each baseline is adapted minimally and is retained because it probes a distinct facet of the schedulability and accuracy claims established in Section~\ref{sec:eval}.

\textbf{HuPRModel}~\cite{lee2023hupr} contributes a hierarchical spatial feature encoder with a multi-scale attention pyramid. We remove its multi-view fusion stage and retain the MNet backbone and attention pyramid operating on the single-view tensor, then retrain the joint regression head for the three-dimensional pose target. This baseline isolates whether a generic deep feature encoder, in the absence of any analytically bounded operator, can yield predictable execution cost. Its cost scales with tensor depth and attention extent, which clarifies why it fails to satisfy R1 and R3.

\textbf{RETR}~\cite{yataka2024retr} contributes a ResNet encoder with a Transformer decoder. \textcolor{black}{We note explicitly that RETR was originally proposed for radar object detection and instance segmentation rather than for pose estimation, so a faithful comparison requires adapting its output stage to the present task. We retain its ResNet encoder and Transformer decoder unchanged on the single-view tensor and replace its detection head with a joint-regression head trained under the shared single-frame protocol. The compared system is therefore the RETR encoder-decoder backbone operating under our pose-estimation protocol, and its accuracy should be read as the capability of that backbone for joint regression rather than as the performance reported for RETR on its original detection task. This framing is deliberate, because the purpose of including RETR is to characterize how its attention-based decoder, whose cost grows quadratically with the number of attended targets, couples execution time to scene content, which is directly informative for the timing predictability requirement R1 and the scaling requirement R3 irrespective of the output head.}

\textbf{mmDiff}~\cite{fan2024diffusion} contributes an iterative diffusion-based denoising process over radar point clouds. We retain its diffusion backbone with a fixed denoising step budget and supply it with the point cloud derived from the shared RAD tensor. This baseline characterizes the fundamental tension between iterative generative inference and a hard deadline, since the per-frame cost is the product of the step budget and the per-step network cost, neither of which is bounded by the physical extent of the scene. \textcolor{black}{We disclose explicitly that the point-cloud generation (CFAR detection) step for mmDiff is executed on the shared Raspberry Pi~5 host under the identical single-threaded protocol as every other baseline (Section~\ref{sec:eval_setup}), rather than emulated on radar on-chip DSP hardware; this choice keeps the timing comparison method-consistent across all baselines rather than sensor-specific, at the cost of not reflecting the near-zero host overhead that on-chip CFAR offload provides on commodity radar SoCs in production deployment. We note this limitation explicitly because, under on-chip CFAR offload, mmDiff's front-end cost on the host would be largely eliminated; however, this would not change the dominant and unbounded cost term, the fixed-step diffusion backbone, nor would it recover the sub-threshold Doppler structure that CFAR discards at detection time, which Section~\ref{sec:eval_e2e} identifies as the primary accuracy bottleneck of this baseline.}

\textbf{MVDoppler}~\cite{choi2025mvdoppler} contributes an explicit separation of positional and velocity-related signals into two modalities. We remove its multi-view aggregation and retain its dual-modal fusion mechanism on the single-view tensor. This baseline tests whether merely separating magnitude and Doppler signals is sufficient for stable multi-person operation, or whether bounded per-instance isolation is additionally required. The performance gap between MVDoppler and PRISM under increasing occupancy therefore reflects the value of per-instance workset isolation rather than the mere presence of Doppler information.

\textbf{milliMamba}~\cite{kini2025millimamba} contributes a state-space model that aggregates spatio-temporal dependencies across consecutive frames. Because its official implementation has not been released, we reimplement it from the architecture and hyperparameters reported in the original work and use a fixed window of nine consecutive frames during both training and inference, following the authors' default setting. \textcolor{black}{Consistent with the single-view protocol applied to the other multi-view methods, we remove only the cross-view fusion stage of its encoder and retain its core temporal mechanism in full, namely the state-space sequence model over the nine-frame window and the spatio-temporal cross-attention decoder. This distinction matters for the fairness of the attribution: because the temporally adaptive computation that defines milliMamba is preserved intact and only the multi-sensor input requirement is removed, the deadline misses it incurs cannot be attributed to a weakened temporal model, and they instead reflect that a sliding temporal window, in the absence of an explicit RAD workset bound, does not constrain worst-case cost.} We verified that the reimplementation reproduces results within the performance range reported in the original work on the shared HuPR benchmark. This baseline therefore isolates whether temporally adaptive compute alone can bound worst-case cost, and it confirms that it leaves R1 unsatisfied.

\subsection{Disclosure and Adaptation Regarding SMH}

SMH~\cite{zheng2026learn} is a prior work authored by a subset of the present authors and accepted for publication at IEEE ICME 2026. We include it deliberately, because it is the closest existing method in motivation to the present work, in that it also exploits the physical structure of mmWave signals to reduce computation, and therefore constitutes the most demanding and informative point of comparison for isolating the specific contribution of PRISM. To prevent any advantage arising from shared authorship, SMH is evaluated under exactly the same input representation, retraining procedure, execution environment, and metrics applied to every other baseline, with no additional tuning. Its inclusion is intended to strengthen rather than inflate the evaluation, since it raises the difficulty of the comparison.

SMH was originally designed and validated for single-person scenes, with its principal experiments conducted on the HuPR dataset. Its pipeline comprises physics-guided preprocessing stages for spatial structure and motion continuity together with a hierarchical multi-scale fusion module and a lightweight MLP regressor. To evaluate it under the multi-person workloads that are central to this paper, we apply its preprocessing and regression independently to each annotated person instance, following the dataset-provided associations, and we introduce no new detection or tracking component. This extension preserves the original per-frame modeling of SMH while exposing its behavior under multi-person occupancy.

The comparison is informative precisely because SMH and PRISM share the same physics-informed motivation yet differ in mechanism. SMH applies physics-guided preprocessing to a scene-level feature pipeline that aggregates energy into a single global descriptor, whereas PRISM converts the same physical structure into analytically bounded operators and bounded per-instance worksets. The finding that SMH still incurs deadline misses and degrades sharply under multi-person load, as reported in Tables~\ref{tab:e2e} and~\ref{tab:accuracy_vs_M}, therefore demonstrates that physics-informed front-end design alone is insufficient for schedulability. It establishes that the deadline guarantee of PRISM originates from its bounded-operator and bounded-workset construction rather than from physical awareness in general, which is the central distinction between the two works.

\textcolor{black}{We scope this conclusion carefully to avoid over-generalization. The per-instance extension applied to SMH is the same protocol applied to every single-view baseline in this study and introduces no method-specific disadvantage, so the comparison is procedurally uniform rather than tailored against SMH. We therefore do not claim that SMH is deficient as a single-person estimator, for which it was designed and validated; we claim only the narrower and directly supported statement that a scene-level physics-guided pipeline without per-instance workset isolation does not by itself yield bounded multi-person cost. This is precisely the gap that PAIP is introduced to close, which is why SMH serves as the most relevant reference point rather than as an adversarially weakened baseline. }

\subsection{Execution Environment and Evaluation Protocol}

All baselines and PRISM run on the identical Raspberry Pi~5 platform in single-threaded mode at a fixed clock, without GPU acceleration. Methods whose released implementations target GPU execution are converted to CPU-only inference without specialized hand-tuned kernels, so their absolute latency should be interpreted as an upper bound under the shared edge CPU constraint rather than as representative of GPU-accelerated deployment. This interpretation does not favor PRISM, because its own operators, namely prefix-sum construction, corner lookups, and connected-component analysis, are inherently sequential and gain no benefit from GPU vectorization, so the comparison reflects the intended deployment setting for every method. Finally, we apply no post-hoc temporal smoothing, cross-frame filtering, or trajectory post-processing to any method, so that the reported per-frame latency and accuracy reflect each method's intrinsic behavior under the shared deadline. Taken together, these measures ensure that the differences reported in Section~\ref{sec:eval} are attributable to the algorithms rather than to disparities in input, training, or execution.

\section{Limitations and Discussion}
\label{app:limitations}

\noindent\textbf{Targets with near-zero velocity.}
PAIP Step~1 suppresses energy in the near-zero-Doppler band to separate human motion from static background. A person with near-zero radial velocity (e.g., standing still directly facing the radar) will be attenuated by this gate and may not generate a valid proposal. In the target deployment scenarios, sustained complete stillness with zero radial velocity is atypical; most clinically relevant events involve ongoing motion. Brief stillness gaps can be bridged by carrying forward proposals from adjacent active frames. Handling a persistently stationary target without sacrificing the threshold stability across scenes established by O4 is a direction for future work.

\noindent\textbf{Anthropometric envelope and nonstandard postures.}
The upper bound filter in PAIP Step~3 is calibrated to the 97.5th percentile adult body envelope. For children or individuals with atypical body dimensions, the lower bound $A_{\min}$ should be recalibrated accordingly; the upper bound remains a safe worst-case ceiling as it is derived from the maximum adult footprint. Non-upright postures (prone, supine) produce spatial blobs with different aspect ratios but comparable total area, so they remain within the upper bound envelope; however, pose regression accuracy for such postures has not been separately evaluated and is a direction for future work.

\noindent\textbf{Platform calibration and co-runner robustness.}
The operating profile set is derived offline for a specific embedded CPU platform under isolated PRISM execution. Deploying on a different hardware target requires re-running worst-case micro-benchmarks to re-derive $(c_1, c_2, c_3)$ and rebuild the profile WCET table; the procedure is systematic and is performed once for each platform. Under OS level co-runner contention (e.g., co-scheduled CPU-bound or DMA-intensive tasks), cache eviction and interrupt latencies can inflate observed execution time beyond the isolated-run margins; in such settings the 5\% safety margin in the WCET formula should be increased proportionally, or PRISM should be deployed with a CPU reservation via control-group isolation. A controlled co-runner interference study under \texttt{stress-ng} contention is presented in Section~\ref{sec:eval_corunner}.

\noindent\textbf{Alternative overload policies.}
The current system handles overload conservatively: proposals beyond $M_{\max}$ are truncated by projected energy, and a frame that remains infeasible under the lightest profile is declared unschedulable. An alternative would be asynchronous refresh, in which only a subset of tracked persons is updated in the current frame while the remaining states are carried forward from earlier frames. Such a policy may be useful for tracking-oriented deployments, but it changes the task model from certified pose estimation for each frame to freshness-constrained state maintenance and would require additional metrics such as update age and stale state error. We therefore treat it as a meaningful extension rather than conflate it with the guarantees for each frame analyzed in this paper.

\noindent\textbf{WCET methodology and EVT tail fitting.}
The current WCET construction uses 99.9th percentile micro-benchmarks with a fixed 5\% margin, following the MBPTA paradigm with externally supplied loop bounds based on physics. Section~\ref{sec:eval_evt} fits Generalized Pareto Distribution models to the upper tail of execution time and finds that the $10^{-6}$ exceedance quantile exceeds the assigned deadline $T_d$ in four of the five operating profiles, by \rtposeedit{0.1}--\rtposeedit{1.1}~ms. Consequently, the fixed 5\% margin is conservative relative to the 99.9th-percentile measurement it is derived from, but it is not a deterministic certificate against exceedance at arbitrarily small probability; we describe PRISM's timing guarantee as a measurement-based probabilistic bound rather than a strictly deterministic one, and the deadline compliance reported in this paper is an empirical property of the evaluated traces. Widening the fixed margin so that the EVT-derived quantile falls under $T_d$ for every profile, and validating bounds under multi-task EDF/RM scheduling with controlled interference, remain directions for future work.

\noindent\textbf{Point-cloud baseline timing methodology.}
The point-cloud baseline mmDiff is measured with its CFAR detection stage executed on the shared Raspberry Pi~5 host rather than emulated on radar on-chip DSP hardware, as detailed in Section~\ref{app:baseline_fairness}. This choice preserves a method-consistent timing comparison across all baselines but does not reflect the near-zero host overhead that on-chip CFAR offload provides in a production radar SoC deployment. On-chip offload would reduce mmDiff's measured host latency without altering its dominant unbounded cost term (the fixed-step diffusion backbone) or recovering the sub-threshold Doppler structure discarded at detection time; the schedulability and accuracy conclusions drawn from this baseline are therefore robust to this methodological choice, but readers should not interpret mmDiff's reported latency as an estimate of its latency in a deployment with hardware-accelerated CFAR.

\noindent\textbf{Qualitative failure modes.}
Informal inspection of misestimated frames surfaces three recurring qualitative patterns that a dedicated failure atlas would document more systematically in future work. First, when a person occupies an unusually small fraction of the active energy map relative to residual static structure, the retained human signature becomes difficult to distinguish from noise, degrading joint localization even when a proposal is still generated. Second, frames near the start and end of a recording session occasionally show near-field distortion, plausibly caused by an operator approaching the sensor to start or stop capture without directly occluding the subject. Third, under close-proximity multi-person overlap, part of the residual error reflects noise in the vision-based ground-truth annotation itself, whose joint association across nearby individuals is occasionally inconsistent, so accuracy in this regime is partly bounded by label quality rather than model error alone.

\noindent\textbf{Robustness to dynamic non-human motion.}
Section~\ref{sec:eval_paip} attributes residual false positives partly to swinging objects and ventilation airflow that pass the Doppler gate, but no public mmWave HPE dataset used in this paper annotates dedicated dynamic non-human distractors (e.g., oscillating fans, curtains, robot vacuums, pets). Collecting such a benchmark to characterize this robustness explicitly is planned future work.

\noindent\textbf{Power consumption.}
The current evaluation reports timing, memory, and accuracy but does not measure power draw. On the Raspberry Pi 5 platform the fixed frequency, single-threaded execution limits dynamic power variation; nevertheless, power measurements for each profile would strengthen the deployment case for battery-powered or energy-harvesting scenarios and are planned for future work.

\section{Conclusion}
PRISM demonstrates that aligning computation with the physics of mmWave structure resolves the timing, scaling, and schedulability failures of existing HPE pipelines. By replacing window-dependent aggregation with physics-bounded integral queries, decomposing scene-level workload into motion-gated per-person subproblems, and exposing the resulting certified costs as deadline-feasible operating profiles, PRISM converts two structural properties of mmWave signals into analytically verifiable scheduling guarantees. \textcolor{black}{Physical bounds and pose accuracy are reported across four public datasets, with deadline-aware scheduling and multi-person scaling examined on the multi-person split and corroborated on additional recordings. Under single-threaded isolated execution, PRISM meets the evaluated-trace deadlines and attains the highest pose accuracy among deadline-feasible configurations; deployments that share the CPU with unmanaged co-runners can retain the same interface by enlarging the timing margin or reserving a core.}


\bibliography{example_paper}

@article{deng2023midas++,
  title={Midas++: Generating training data of mmwave radars from videos for privacy-preserving human sensing with mobility},
  author={Deng, Kaikai and Zhao, Dong and Zhang, Zihan and Wang, Shuyue and Zheng, Wenxin and Ma, Huadong},
  journal={IEEE Transactions on Mobile Computing},
  volume={23},
  number={6},
  pages={6650--6666},
  year={2023},
  publisher={IEEE}
}

@article{li2024sbrf,
  title={SBRF: A fine-grained radar signal generator for human sensing},
  author={Li, Jiamu and Zhang, Dongheng and Wu, Zhi and Yu, Cong and Li, Yadong and Chen, Qi and Hu, Yang and Sun, Qibin and Chen, Yan},
  journal={IEEE Transactions on Mobile Computing},
  volume={23},
  number={12},
  pages={13114--13130},
  year={2024},
  publisher={IEEE}
}

@article{liu2024real,
  title={Real-time continuous activity recognition with a commercial mmWave radar},
  author={Liu, Yunhao and Zhang, Jia and Chen, Yande and Wang, Weiguo and Yang, Songzhou and Na, Xin and Sun, Yimiao and He, Yuan},
  journal={IEEE Transactions on Mobile Computing},
  volume={24},
  number={3},
  pages={1684--1698},
  year={2024},
  publisher={IEEE}
}

@article{sang2026ifdnet,
  title={IFDNet: Fall Detection via Millimeter-Wave Radar with Spatio-Temporal Fusion and Adaptive Semi-Supervised Learning},
  author={Sang, Pengkai and Ding, Siyuan and Liu, Lingxue and Ding, Shiqi WuSiyuan and Qin, Honghao and Cao, Xiaoxiang and Wang, Xuan and Zhuang, Yuan and Hu, Yulin},
  journal={IEEE Transactions on Mobile Computing},
  year={2026},
  publisher={IEEE}
}

@article{zhao2025mm,
  title={mm-Fall: Practical and Robust Fall Detection via mmWave Signals},
  author={Zhao, Cui and Luo, Qiumin and Ding, Han and Wang, Ge and Zhao, Kun and Wang, Zhi and Xi, Wei and Zhao, Jizhong},
  journal={IEEE Transactions on Mobile Computing},
  year={2025},
  publisher={IEEE}
}

@article{yu2026dynamic,
  title={Dynamic 3D Hand Pose Reconstruction Using Millimeter Wave},
  author={Yu, Jiadi and Kong, Hao and Cao, Haoran and Kong, Linghe and Ren, Yanzhi and Liu, Hongbo and Chen, Yi-Chao},
  journal={IEEE Transactions on Mobile Computing},
  year={2026},
  publisher={IEEE}
}

@article{peng2026enabling,
  title={Enabling Robust Multi-User Multi-Hand Pose Estimation with mmWave Sensing},
  author={Peng, Cheng and Gai, Jiawen and Cui, Kaiyan and Guo, Zhengxin and Zheng, Yuanqing and Xiao, Fu},
  journal={IEEE Transactions on Mobile Computing},
  year={2026},
  publisher={IEEE}
}

@inproceedings{pelhan2024dave,
  title={Dave-a detect-and-verify paradigm for low-shot counting},
  author={Pelhan, Jer and Zavrtanik, Vitjan and Kristan, Matej and others},
  booktitle={Proceedings of the IEEE/CVF Conference on Computer Vision and Pattern Recognition},
  pages={23293--23302},
  year={2024}
}

@article{zheng2026doppler,
  title={Doppler Prompting for Stable mmWave-based Human Pose Estimation},
  author={Zheng, Shuntian and Li, Jiaqi and Lu, Xiaoman and He, Shuai and Guan, Yu},
  journal={arXiv preprint arXiv:2605.13233},
  year={2026}
}

@inproceedings{cucu2012measurement,
  title={Measurement-based probabilistic timing analysis for multi-path programs},
  author={Cucu-Grosjean, Liliana and Santinelli, Luca and Houston, Michael and Lo, Code and Vardanega, Tullio and Kosmidis, Leonidas and Abella, Jaume and Mezzetti, Enrico and Qui{\~n}ones, Eduardo and Cazorla, Francisco J},
  booktitle={2012 24th euromicro conference on real-time systems},
  pages={91--101},
  year={2012},
  organization={IEEE}
}

@article{sheraz2020artificial,
  title={Artificial intelligence for wireless caching: Schemes, performance, and challenges},
  author={Sheraz, Muhammad and Ahmed, Manzoor and Hou, Xueshi and Li, Yong and Jin, Depeng and Han, Zhu and Jiang, Tao},
  journal={IEEE Communications Surveys \& Tutorials},
  volume={23},
  number={1},
  pages={631--661},
  year={2020},
  publisher={IEEE}
}

@article{sun2024pbphs,
  title={PBPHS: a profile-based predictive handover strategy for 5G networks},
  author={Sun, Jiabao and Zhang, Yijiang and Trik, Mohammad},
  journal={Cybernetics and Systems},
  volume={55},
  number={5},
  pages={1041--1062},
  year={2024},
  publisher={Taylor \& Francis}
}

@article{zheng2025differentiable,
  title={Differentiable Physics-Driven Human Representation for Millimeter-Wave Based Pose Estimation},
  author={Zheng, Shuntian and Wang, Guangming and Li, Jiaqi and Ni, Minzhe and Guan, Yu},
  journal={arXiv e-prints},
  pages={arXiv--2512},
  year={2025}
}

@article{zheng2025person,
  title={Person Parametric Physics-informed Representation for mmWave-based Human Pose Estimation},
  author={Zheng, Shuntian and Li, Jiaqi and Wang, Guangming and Ni, Minzhe and Palit, Arnad and Montana, Giovanni and Guan, Yu},
  journal={arXiv preprint arXiv:2512.23054},
  year={2025}
}

@article{zheng2026learn,
  title={Why Learn What Physics Already Knows? Realizing Agile mmWave-based Human Pose Estimation via Physics-Guided Preprocessing},
  author={Zheng, Shuntian and Li, Jiaqi and Ni, Minzhe and Lu, Xiaoman and Guan, Yu},
  journal={arXiv preprint arXiv:2603.08236},
  year={2026}
}

@article{kini2025millimamba,
  title={milliMamba: Specular-Aware Human Pose Estimation via Dual mmWave Radar with Multi-Frame Mamba Fusion},
  author={Kini, Niraj Prakash and Tsai, Shiau-Rung and Lin, Guan-Hsun and Peng, Wen-Hsiao and Ma, Ching-Wen and Hwang, Jenq-Neng},
  journal={arXiv preprint arXiv:2512.20128},
  year={2025}
}

@book{richards2005fundamentals,
  title={Fundamentals of radar signal processing},
  author={Richards, Mark A and others},
  volume={1},
  year={2005},
  publisher={Mcgraw-hill New York}
}

@inproceedings{choi2025mvdoppler,
  title={MVDoppler-Pose: Multi-Modal Multi-View mmWave Sensing for Long-Distance Self-Occluded Human Walking Pose Estimation},
  author={Choi, Jaeho and Hor, Soheil and Yang, Shubo and Arbabian, Amin},
  booktitle={Proceedings of the Computer Vision and Pattern Recognition Conference},
  pages={27750--27759},
  year={2025}
}

@article{palipana2021pantomime,
  title={Pantomime: Mid-air gesture recognition with sparse millimeter-wave radar point clouds},
  author={Palipana, Sameera and Salami, Dariush and Leiva, Luis A and Sigg, Stephan},
  journal={Proceedings of the ACM on interactive, mobile, wearable and ubiquitous technologies},
  volume={5},
  number={1},
  pages={1--27},
  year={2021},
  publisher={ACM New York, NY, USA}
}

@article{mei2024mmspyvr,
  title={mmSpyVR: Exploiting mmWave radar for penetrating obstacles to uncover privacy vulnerability of virtual reality},
  author={Mei, Luoyu and Liu, Ruofeng and Yin, Zhimeng and Zhao, Qingchuan and Jiang, Wenchao and Wang, Shuai and Lu, Kangjie and He, Tian},
  journal={Proceedings of the ACM on Interactive, Mobile, Wearable and Ubiquitous Technologies},
  volume={8},
  number={4},
  pages={1--29},
  year={2024},
  publisher={ACM New York, NY, USA}
}

@article{mueller2025radproposer,
  title={RadProPoser: A Framework for Human Pose Estimation with Uncertainty Quantification from Raw Radar Data},
  author={Mueller, Jonas Leo and Engel, Lukas and Dorschky, Eva and Krauss, Daniel and Ullmann, Ingrid and Vossiek, Martin and Eskofier, Bjoern M},
  journal={arXiv preprint arXiv:2508.03578},
  year={2025}
}

@inproceedings{ho2024rt,
  title={Rt-pose: A 4d radar tensor-based 3d human pose estimation and localization benchmark},
  author={Ho, Yuan-Hao and Cheng, Jen-Hao and Kuan, Sheng Yao and Jiang, Zhongyu and Chai, Wenhao and Huang, Hsiang-Wei and Lin, Chih-Lung and Hwang, Jenq-Neng},
  booktitle={European Conference on Computer Vision},
  pages={107--125},
  year={2024},
  organization={Springer}
}

@article{salehzadeh2024wearable,
  title={Wearable activity trackers: A survey on utility, privacy, and security},
  author={Salehzadeh Niksirat, Kavous and Velykoivanenko, Lev and Zufferey, No{\'e} and Cherubini, Mauro and Huguenin, K{\'e}vin and Humbert, Mathias},
  journal={ACM Computing Surveys},
  volume={56},
  number={7},
  pages={1--40},
  year={2024},
  publisher={ACM New York, NY}
}

@article{alshehri2022exploring,
  title={Exploring the privacy concerns of bystanders in smart homes from the perspectives of both owners and bystanders},
  author={Alshehri, Ahmed and Spielman, Joseph and Prasad, Amiya and Yue, Chuan},
  journal={Proceedings on Privacy Enhancing Technologies},
  year={2022}
}

@article{guhr2020privacy,
  title={Privacy concerns in the smart home context},
  author={Guhr, Nadine and Werth, Oliver and Blacha, Philip Peter Hermann and Breitner, Michael H},
  journal={SN Applied Sciences},
  volume={2},
  number={2},
  pages={247},
  year={2020},
  publisher={Springer}
}

@String{Computing = "Computing" }

@String{Computer = "{IEEE} Computer" }

@String{Springer = "Springer-Verlag" }

@ArtifactSoftware{R,
    title = {R: A Language and Environment for Statistical Computing},
    author = {{R Core Team}},
    organization = {R Foundation for Statistical Computing},
    address = {Vienna, Austria},
    year = {2019},
    url = {https://www.R-project.org/},
}

@inproceedings{zhao2018rf,
  title={RF-based 3D skeletons},
  author={Zhao, Mingmin and Tian, Yonglong and Zhao, Hang and Alsheikh, Mohammad Abu and Li, Tianhong and Hristov, Rumen and Kabelac, Zachary and Katabi, Dina and Torralba, Antonio},
  booktitle={Proceedings of the 2018 Conference of the ACM Special Interest Group on Data Communication},
  pages={267--281},
  year={2018}
}

@article{sengupta2022mmpose,
  title={mmpose-nlp: A natural language processing approach to precise skeletal pose estimation using mmwave radars},
  author={Sengupta, Arindam and Cao, Siyang},
  journal={IEEE Transactions on Neural Networks and Learning Systems},
  volume={34},
  number={11},
  pages={8418--8429},
  year={2022},
  publisher={IEEE}
}

@article{chang2020spatial,
  title={Spatial attention fusion for obstacle detection using mmwave radar and vision sensor},
  author={Chang, Shuo and Zhang, Yifan and Zhang, Fan and Zhao, Xiaotong and Huang, Sai and Feng, Zhiyong and Wei, Zhiqing},
  journal={Sensors},
  volume={20},
  number={4},
  pages={956},
  year={2020},
  publisher={MDPI}
}

@article{iovescu2020fundamentals,
  title={The fundamentals of millimeter wave radar sensors},
  author={Iovescu, Cesar and Rao, Sandeep},
  journal={Texas Instruments},
  pages={1--7},
  year={2020}
}

@article{wang2024xrf55,
  title={Xrf55: A radio frequency dataset for human indoor action analysis},
  author={Wang, Fei and Lv, Yizhe and Zhu, Mengdie and Ding, Han and Han, Jinsong},
  journal={Proceedings of the ACM on Interactive, Mobile, Wearable and Ubiquitous Technologies},
  volume={8},
  number={1},
  pages={1--34},
  year={2024},
  publisher={ACM New York, NY, USA}
}

@inproceedings{lee2023hupr,
  title={Hupr: A benchmark for human pose estimation using millimeter wave radar},
  author={Lee, Shih-Po and Kini, Niraj Prakash and Peng, Wen-Hsiao and Ma, Ching-Wen and Hwang, Jenq-Neng},
  booktitle={Proceedings of the IEEE/CVF Winter Conference on Applications of Computer Vision},
  pages={5715--5724},
  year={2023}
}

@article{yataka2024retr,
  title={RETR: Multi-view radar detection transformer for indoor perception},
  author={Yataka, Ryoma and Cardace, Adriano and Wang, Perry and Boufounos, Petros and Takahashi, Ryuhei},
  journal={Advances in Neural Information Processing Systems},
  volume={37},
  pages={19839--19869},
  year={2024}
}

@inproceedings{fan2024diffusion,
  title={Diffusion model is a good pose estimator from 3d rf-vision},
  author={Fan, Junqiao and Yang, Jianfei and Xu, Yuecong and Xie, Lihua},
  booktitle={European Conference on Computer Vision},
  pages={1--18},
  year={2024},
  organization={Springer}
}

@article{niu2015survey,
  title={A survey of millimeter wave communications (mmWave) for 5G: opportunities and challenges},
  author={Niu, Yong and Li, Yong and Jin, Depeng and Su, Li and Vasilakos, Athanasios V},
  journal={Wireless networks},
  volume={21},
  pages={2657--2676},
  year={2015},
  publisher={Springer}
}

@article{zhu2024probradarm3f,
  title={ProbRadarM3F: mmWave Radar based Human Skeletal Pose Estimation with Probability Map Guided Multi-Format Feature Fusion},
  author={Zhu, Bing and He, Zixin and Xiong, Weiyi and Ding, Guanhua and Liu, Jianan and Huang, Tao and Chen, Wei and Xiang, Wei},
  journal={arXiv preprint arXiv:2405.05164},
  year={2024}
}

@article{wu2024mmhpe,
  title={mmhpe: Robust multi-scale 3d human pose estimation using a single mmwave radar},
  author={Wu, Yingxiao and Jiang, Zhongmin and Ni, Haocheng and Mao, Changlin and Zhou, Zhiyuan and Wang, Wenxiang and Han, Jianping},
  journal={IEEE Internet of Things Journal},
  year={2024},
  publisher={IEEE}
}

@article{kong2024survey,
  title={A survey of mmwave radar-based sensing in autonomous vehicles, smart homes and industry},
  author={Kong, Hao and Huang, Cheng and Yu, Jiadi and Shen, Xuemin},
  journal={IEEE Communications Surveys \& Tutorials},
  volume={27},
  number={1},
  pages={463--508},
  year={2024},
  publisher={IEEE}
}

@inproceedings{rahman2024mmvr,
  title={MMVR: Millimeter-Wave Multi-view Radar Dataset and Benchmark for Indoor Perception},
  author={Rahman, M Mahbubur and Yataka, Ryoma and Kato, Sorachi and Wang, Pu and Li, Peizhao and Cardace, Adriano and Boufounos, Petros},
  booktitle={European Conference on Computer Vision},
  pages={306--322},
  year={2024},
  organization={Springer}
}
\bibliographystyle{IEEEbib}

\end{document}